\documentclass[fleqn,usenatbib]{mnras}

\usepackage{newtxtext,newtxmath}

\usepackage[T1]{fontenc}

\DeclareRobustCommand{\VAN}[3]{#2}
\let\VANthebibliography\thebibliography
\def\thebibliography{\DeclareRobustCommand{\VAN}[3]{##3}\VANthebibliography}

\usepackage{graphicx}	
\usepackage{amsmath}	
\usepackage{amssymb}	

\usepackage{ulem}

\title[Kinetically Coupled Scalar Fields]{Kinetically Coupled Scalar Fields Model and Cosmological Tensions}

\author[Gang Liu et al.]{
Gang Liu,\thanks{E-mail: liugang\_dlut@mail.dlut.edu.cn}
Zhihuan Zhou,
Yuhao Mu
and Lixin Xu\thanks{E-mail: lxxu@dlut.edu.cn}
\\
Institute of Theoretical Physics\\
School of Physics\\
Dalian University of Technology\\
Dalian 116024, People's Republic of China
}

\date{Accepted XXX. Received YYY; in original form ZZZ}

\pubyear{2023}

\begin{document}
\label{firstpage}
\pagerange{\pageref{firstpage}--\pageref{lastpage}}
\maketitle

\begin{abstract}
In this paper, we investigate the kinetically coupled early dark energy (EDE) and 
scalar field dark matter to address cosmological tensions. The EDE 
model presents an intriguing theoretical approach to resolving the Hubble tension, but 
it exacerbates the large-scale structure 
tension. We consider the interaction between dark matter and EDE, such that 
the drag of dark 
energy on dark matter suppresses structure growth, which can alleviate large-scale 
structure tension. We replace cold dark matter with scalar field dark 
matter, which has the property of suppressing structure growth on small scales. 
We employed the Markov Chain Monte Carlo method to constrain the model parameters, 
our new model 
reveals a non-zero coupling constant of $0.030 \pm 0.026$ at a 68\% confidence 
level. The coupled model yields the Hubble constant value of $72.38^{+0.71}_{-0.82}$
\,km\,/\,s\,/\,Mpc, 
which resolves the Hubble tension. However, similar to the EDE model, it also obtains 
a larger $S_8$ value compared to the $\Lambda$CDM model, further exacerbating the 
large-scale structure tension. 
The EDE model and the new model yield the best-fit values of $0.8316$ and 
$0.8146$ for $S_8$, respectively, indicating that the new model partially alleviates the 
negative effect of the EDE model. However, this signature disappears when comparing 
marginalised posterior probabilities, and both models produce similar results. The values 
obtained from the EDE model and the new model are $0.822^{+0.011}_{-0.0093}$ and 
$0.819^{+0.013}_{-0.0092}$, respectively, at a 68\% confidence level.
\end{abstract}




\section{Introduction}

The standard $\Lambda$CDM model established over the past decades has received 
significant support from cosmological data. However, with the continuous improvement 
of experimental measurements, inconsistencies between some observational results and 
the predicted results of the $\Lambda$CDM model have become apparent. The most 
well-known inconsistencies are the Hubble tension and the large-scale structure 
tension. The Hubble tension refers to the discrepancy in the value of the Hubble 
constant derived from early universe observations based on the $\Lambda$CDM model and 
late-time measurements that are independent of the model \citet{Verde_2019}. 
Specifically, the best-fit 
of the $\Lambda$CDM model to the \textit{Planck} 2018 cosmic microwave background 
(CMB) data gives $67.37\pm0.54$\,km\,/\,s\,/\,Mpc \citet{planck2020}, whereas the SH0ES 
collaboration, which uses the distance ladder method at low redshifts, gives 
$73.04\pm1.04$\,km\,/\,s\,/\,Mpc \citet{Riess_2022}, with a statistical error of 4.8$\sigma$ 
between them.

Another relatively mild tension refers to the inconsistency between 
the amplitude of density fluctuations $\sigma_8$ measured from large-scale structure 
observations and that measured from CMB observations \citet{PRL.111.161301, Hildebrandt_2020}. 
The quantity 
$S_8\equiv\sigma_8\sqrt{(\Omega_m/0.3)}$ is commonly defined to characterise 
this tension, where $\Omega_m$ denotes the matter energy density fraction. 
The \textit{Planck} 2018 best-fit $\Lambda$CDM model yields a value of $S_8$ 
equal to $0.834\pm0.016$ \citet{planck2020}. However, large-scale structure surveys 
such as the Dark Energy Survey Year-3 (DES-Y3) have reported a smaller value 
$0.776\pm0.017$ \citet{PRD.105.023520}.

Currently, it is unclear whether these two tensions stem from unknown statistical 
errors or inappropriate calibrations \citet{Blanchard_2021, Mortsell_2022}, 
for instance, the potential presence of crowding effects in the 
Cepheid photometry \citet{Freedman:2023zdo}, as well as the inaccuracy in the colour/dust 
correction of type Ia supernovae within calibration galaxies \citet{stac1878}; 
or herald new physics beyond the standard model. 
In any case, it is still meaningful to explore alternative models to the $\Lambda$CDM 
model and investigate dark matter, dark energy, or dark radiation models that can 
alleviate these tensions.

Numerous approaches have been proposed to address the Hubble tension, which can be 
mainly classified into two categories: modifying the physics of the late Universe 
\citet{Guo_2019, Li_2019, Zhouzh, liu2023cosmological} 
and introducing new physics prior to recombination 
\citet{PRD.94.103523, PRD.98.083525, ALEXANDER2019134830, PRD.100.063542, 
PRD.101.083537, PRD.102.123544, Liu:2023haw,Liu:2023rvo}. 
However, these methods are confronted with various challenges. For instance, the 
late-time models are competitively constrained by independent observations at low redshift 
and generally cannot account for SH0ES measurements 
\citet{PRD.101.103517, PRD.101.083524, stab1070}. 
Early-time models that achieve relative success in increasing $H_0$ often exacerbate 
the tension in large-scale structure \citet{PRD.102.043507, PRD.106.043525}.
Despite this, the relative success in relieving Hubble tension through modifications 
to early universe has stimulated further investigation into such models.
In this paper, we will focus on one of the most absorbing cases in early models, 
namely early dark energy (EDE) model 
\citet{PRD.94.103523, PRD.100.063542, PRL.122.221301}, and address its associated issues.

EDE is composed of an ultra-light scalar field, which only make a significant 
contribution during the epoch approaching matter-radiation equality, and is negligible 
during other epochs. The addition of a new component decreases the sound horizon at 
recombination, thereby allowing an increase in the value of $H_0$ while keeping the 
sound horizon angular scale unchanged.

The phenomenological parameter $z_\mathrm{c}$ is generally used to represent the 
redshift at which the energy density of EDE reaches its peak, and $f_\mathrm{EDE}$ is 
used to denote the ratio of EDE energy density to the total energy density at this 
redshift. The value of $f_\mathrm{EDE}$ reaches approximately 10\% can resolve 
the Hubble tension.

However, obtaining a larger value of $H_0$ through the EDE scenario comes at the cost 
of inducing changes in other cosmological parameters, such as the density of dark 
matter $\omega_\mathrm{c}$, the scalar spectral index $n_\mathrm{s}$, and the amplitude 
of density perturbations $\sigma_8$ \citet{PRD.106.043525}. 
This exacerbates the tension between the model 
and the large-scale structure data. 

One natural idea to address these issues is to introduce the interaction between 
EDE and dark matter. The drag of dark energy on dark matter can inhibit structure 
growth, alleviate the large-scale structure tension exacerbated by EDE.

Previous studies have investigated the interaction between EDE and dark matter, 
such as utilising the Swampland Distance Conjecture \citet{Ooguri_2007, Klaewer_2017, 
Palti} to consider the exponential dependence of the EDE scalar on 
the mass of dark matter \citet{PRD.106.043525, PRD.107.103523, liu2023alleviating}. 
In this work, inspired by string theory, we consider the kinetic coupling between 
EDE and scalar field dark matter (SFDM). 

\begin{figure}
	\includegraphics[width=\columnwidth]{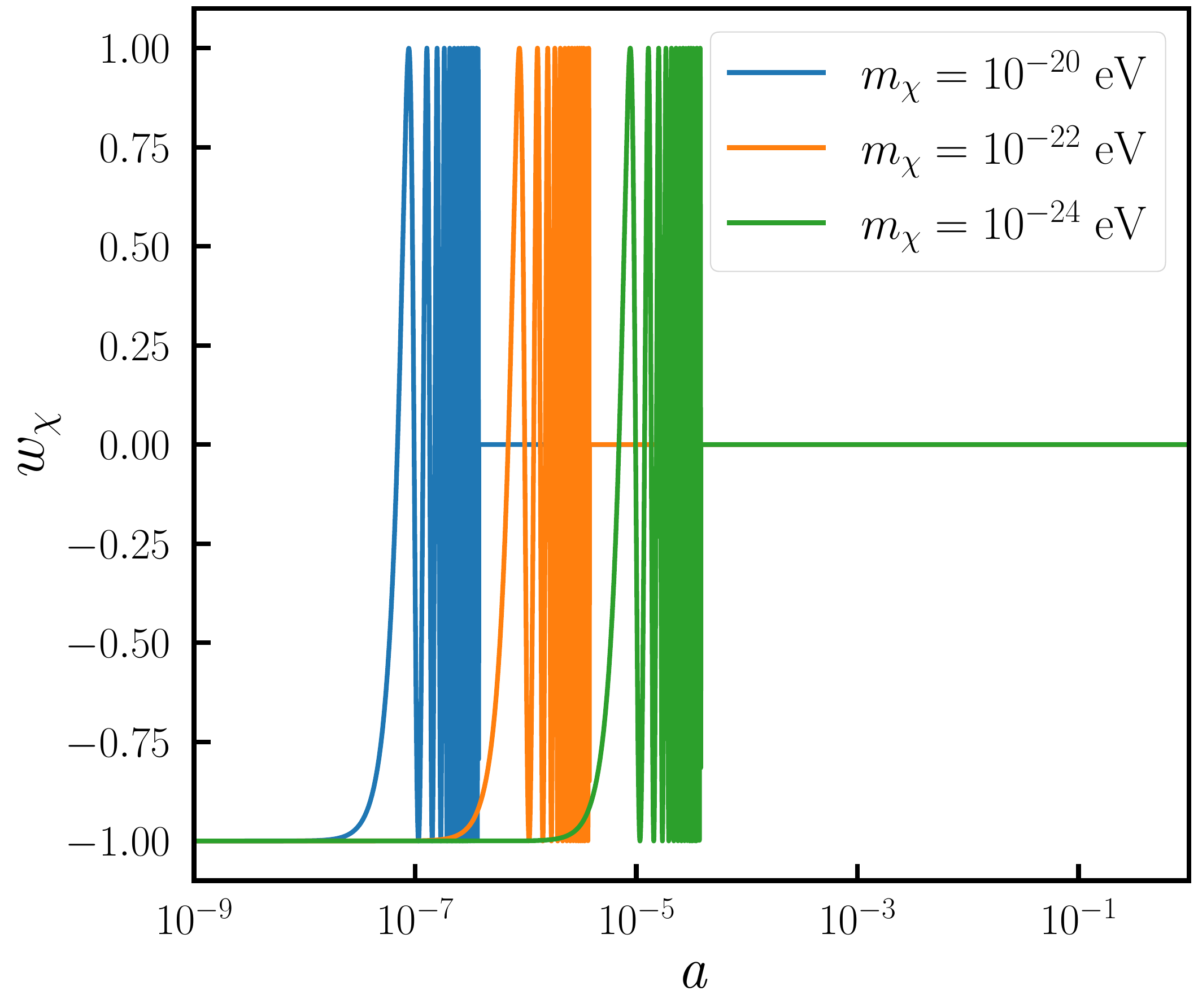}
    \caption{
		The evolution of the equation of state of scalar field dark matter with respect 
        to the scale factor $a$ is investigated. The evolution of SFDM exhibits behavior 
        similar to a cosmological constant in the early universe, then experiences 
        oscillations, and finally evolves similarly to that of cold dark matter.}
    \label{fig:1}
\end{figure}

SFDM, consisting of a light scalar field with a mass of approximately $10^{-22}$ eV, 
is a viable alternative to cold dark matter (CDM) \citet{Ferreira_2021, 
PRD.106.123501}. SFDM forms condensates on small scales,
which suppress structure growth, while on large scales it exhibits behavior consistent 
with that of CDM. Fig.~\ref{fig:1} illustrates the equation of state evolution with respect to 
scale factor for SFDM of varying masses. 
The remaining cosmological parameters adopt the results of the \textit{Planck} 
2018 best-fitting $\Lambda$CDM model. 
In the early universe, SFDM behaves like a 
cosmological constant, then undergoes oscillations before ultimately evolving similarly 
to CDM, and the mass of the scalar field affects the onset time of the oscillation 
process.

The kinetic coupling between two scalar fields has been previously investigated, 
particularly in relation to the axio-dilaton models \citet{ALEXANDER2019134830, 
Alexander_2023}. In this paper, we introduce a EDE scalar-dependent function to the 
kinetic term of SFDM, allowing for energy exchange between the two scalar fields. 
The specific Lagrangian is written as 
\begin{equation}
    \mathcal{L}_\mathrm{int}=-\frac{1}{2}g^{\mu\nu}\partial_{\mu}\phi\partial_{\nu}\phi-V(\phi)
    -\frac{1}{2}\zeta(\phi)g^{\mu\nu}\partial_{\mu}\chi\partial_{\nu}\chi-V(\chi), 
	\label{eq:lag}
\end{equation}
where $\phi$ and $\chi$ represent the EDE scalar and SFDM, respectively. The kinetic 
term of SFDM is multiplied by the coupling function $\zeta(\phi)$.

We conducted a detailed study of the evolution equations for the kinetically 
coupled scalar fields (KCS) model, including both the background and perturbation parts. 
We incorporated various commonly used cosmological data, including the \textit{Planck} 
2018 CMB temperature, polarization, and lensing measurements \cite{osti_1676388,osti_1775409, planck2020}; 
BAO measurements from BOSS DR12, 6dF Galaxy Survey, and SDSS DR7 
\cite{19250.x,stv154, Alam_2017,Buen_Abad_2018}; and the Pantheon supernova Ia sample 
\cite{Scolnic_2018}. Additionally, we utilised the measurement of the Hubble 
constant from SH0ES \cite{Riess_2022} and the results of $S_8$ from the Dark Energy 
Survey Year-3 data \cite{PRD.105.023520}. 

By employing the Markov Chain Monte Carlo (MCMC) method, we constrained the model 
parameters. Utilising the entire dataset, the results revealed non-zero coupling 
constants at a 68\% confidence level of $0.030 \pm 0.026$. The new model 
yielded the Hubble constant of $72.38^{+0.71}_{-0.82}$\,km\,/\,s\,/\,Mpc, indicating its 
capability in addressing the Hubble tension. Similar to the EDE model, the coupling 
model still obtained a larger $S_8$ value compared to the $\Lambda$CDM model, with 
the best-fit value being $0.8146$. 
This value is slightly smaller than the result obtained from the EDE model, 
which is $0.8316$. However, 
when comparing the marginalised posterior probabilities, the results from both models are close. 
The constrained values of $S_8$ obtained from the EDE model and the coupled model are 
$0.822^{+0.011}_{-0.0093}$ and $0.819^{+0.013}_{-0.0092}$, respectively, at a 68\% confidence level.

The structure of this paper is as follows: Section~\ref{sec:kcs} introduces the 
KCS model, discussing its dynamics in both the background and perturbation levels, 
as well as its modifications to the original EDE model. In Section~\ref{sec:nr}, 
we present the numerical results of the new model, including its impact on the 
evolution of the Hubble parameter and 
large-scale structure. Section~\ref{sec:dm} provides an introduction to 
the various cosmological data used in the MCMC analysis, along with the resulting 
parameter constraints. Finally, we summarise our findings in Section~\ref{sec:con}.

\section{Kinetically Coupled Scalar Fields}
\label{sec:kcs}
Considering the Lagrangian shown in equation~(\ref{eq:lag}), we employed the 
potential form of EDE as follows \citet{PRD.102.043507, PRD.101.063523}, 
\begin{equation}
    V(\phi)=m_{\phi}^2f_{\phi}^2[1-\cos(\phi/f_{\phi})]^3+V_{\Lambda},
\end{equation}
where $m_{\phi}$ represents the axion mass, $f_{\phi}$ denotes its decay constant, 
and $V_{\Lambda}$ plays the role of a cosmological constant. For SFDM, we assume a 
simple quadratic potential, 
\begin{equation}
	V(\chi)=\frac{1}{2}m_{\chi}^2\chi^2, 
\end{equation}
with $m_{\chi}$ representing the mass of the SFDM. As for the coupling function, 
a natural choice is to assume an exponential form,
\begin{equation}
	\zeta(\phi)=\exp(\frac{\lambda\phi}{M_{pl}}),
\end{equation}
where $\lambda$ represents a dimensionless constant that characterises the strength 
of the interaction, and $M_{pl}$ is the reduced Planck mass.

\subsection{Background Equations}
The equations of motion for the two scalar fields in a flat Friedmann-Robertson-Walker 
(FRW) metric can be expressed as
\begin{subequations}
    \begin{align}
        &\ddot{\phi}+3H\dot{\phi}+V_{\phi}-\frac{1}{2}\frac{\lambda}{M_{pl}}\exp(\frac{\lambda\phi}{M_{pl}})\dot{\chi}^2=0,\\
        &\ddot{\chi}+(3H+\frac{\lambda\dot{\phi}}{M_{pl}})\dot{\chi}+\exp(-\frac{\lambda\phi}{M_{pl}})m_{\chi}^2\chi=0,
    \end{align} 
	\label{eq:kg}
\end{subequations}
where the dot denotes the derivative with respect to cosmic time, $H$ is the Hubble 
parameter, and $V_{\phi}$ represents the partial derivative of the EDE 
potential with respect to $\phi$. 
It is easy to see that for a non-trivial kinetic coupling function $\zeta(\phi)$, 
the EDE field has a source term proportional to $\dot{\chi}^2$, while the SFDM field 
has a friction term proportional to $\dot{\phi}$. Therefore, if $\dot{\chi}$ is not 
equal to zero and the coupling constant $\lambda$ is positive, energy will be 
transferred from SFDM to EDE.

The energy density and pressure of SFDM can be expressed as 
\begin{subequations}
    \begin{align}
        &\rho_{\chi}=\frac{1}{2}\zeta\dot{\chi}^2+\frac{1}{2}m_{\chi}^2\chi^2,\\
        &p_{\chi}=\frac{1}{2}\zeta\dot{\chi}^2-\frac{1}{2}m_{\chi}^2\chi^2,
    \end{align} 
	\label{eq:rp}
\end{subequations}
where we abbreviate $\zeta(\phi)$ as $\zeta$.
It is convenient to introduce new variables to calculate the motion equation of SFDM 
\citet{PRD.57.4686, Ure_a_L_pez_2016, PRD.96.061301}, 
\begin{subequations}
    \begin{align}
        &\sqrt{\Omega_{\chi}}\sin{\frac{\theta}{2}}=\frac{\sqrt{\zeta}\dot{\chi}}{\sqrt{6}M_{pl}H},\\
        &\sqrt{\Omega_{\chi}}\cos{\frac{\theta}{2}}=-\frac{m_{\chi}\chi}{\sqrt{6}M_{pl}H},\\
        &y_1=\frac{2m_{\chi}}{H},
    \end{align} 
\end{subequations}
where $\Omega_{\chi}=\frac{\rho_{\chi}}{3M_{pl}^2H^2}$ represents the fraction of 
the dark matter density. 
Combining with the Friedmann Equation, 
\begin{equation}
    \frac{\dot{H}}{H^2}=-\frac{3}{2}(1+w_t), 
\end{equation}
where $w_t$ represents the total equation of state, defined as the ratio of the total 
pressure to the total energy density, 
the evolution equation for the new variable is thus 
formulated as, 
\begin{subequations}
    \begin{align}
        &\frac{\dot{\Omega}_{\chi}}{\Omega_{\chi}}=3H(w_t+\cos\theta)+\frac{\lambda\dot{\phi}}{2M_{pl}}(\cos\theta-1),\\
        &\dot{\theta}=Hy_1\exp(-\frac{\lambda\phi}{2M_{pl}})-(3H+\frac{\lambda\dot{\phi}}{2M_{pl}})\sin\theta,\\
        &\dot{y_1}=\frac{3}{2}H(1+w_t)y_1. 
    \end{align} 
\end{subequations}
If we consider the energy density and pressure of the EDE scalar field, 
\begin{subequations}
    \begin{align}
        &\rho_{\phi}=\frac{1}{2}\dot{\phi}^2+V(\phi),\\
        &p_{\phi}=\frac{1}{2}\dot{\phi}^2-V(\phi),
    \end{align} 
\end{subequations}
combined with equation~(\ref{eq:rp}) and using the equations of motion for both 
fields, we obtain the following continuity equations, 
\begin{subequations}
    \begin{align}
        &\dot{\rho_{\chi}}=-3H(\rho_{\chi}+p_{\chi})-\frac{\lambda\dot{\phi}}{2M_{pl}}\rho_{\chi}(1-\cos\theta),\\
        &\dot{\rho_{\phi}}=-3H(\rho_{\phi}+p_{\phi})+\frac{\lambda\dot{\phi}}{2M_{pl}}\rho_{\chi}(1-\cos\theta).
    \end{align}
\end{subequations}
Many coupled dark energy models exhibit this common structure 
\cite{PhysRevD.77.063513, Kazuya_Koyama_2009, Mukherjee2016InSO, Wang_2016, PhysRevD.96.123508}, 
which guarantees that the total stress tensor is covariantly conserved.

\begin{figure*}
	\includegraphics[width=\linewidth]{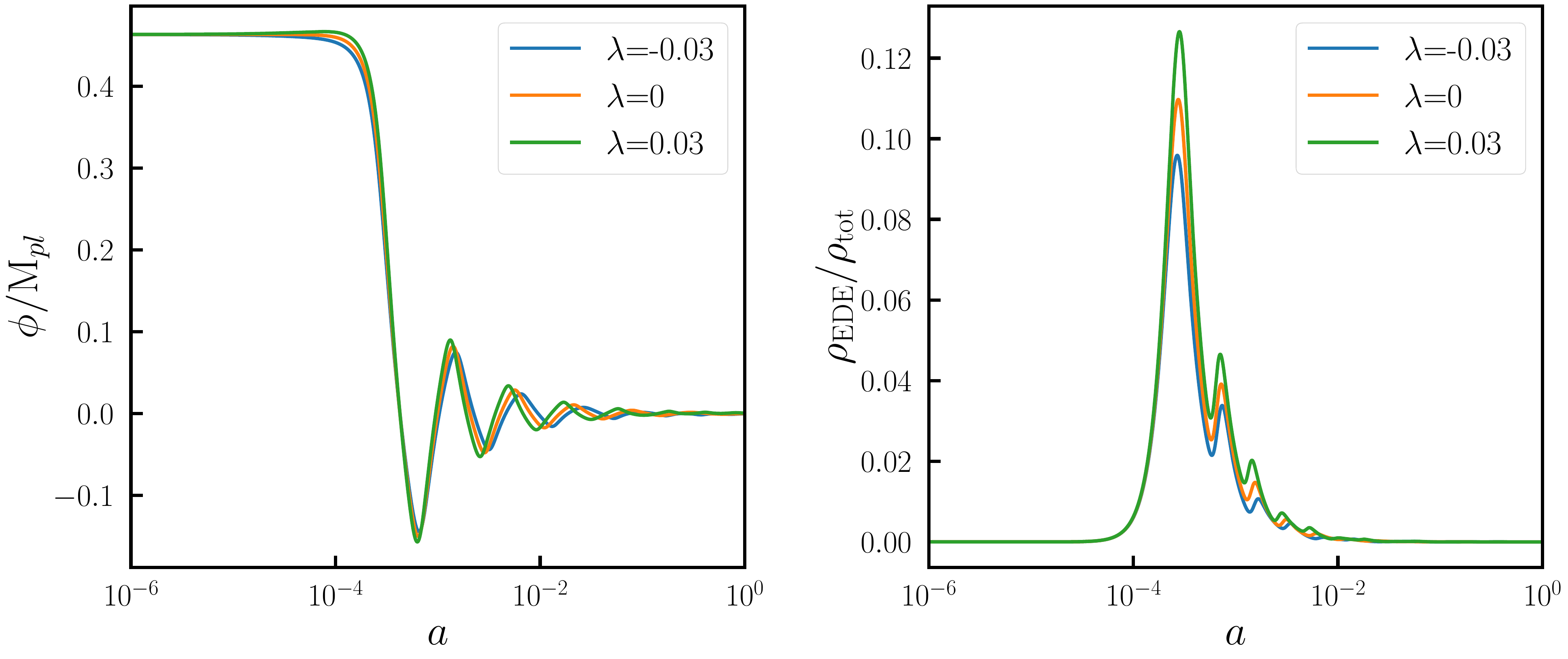}
    \caption{
		The evolution of the EDE scalar (left panel) and the EDE energy density fraction 
        (right panel) with respect to the scale factor is presented. The amplitude and 
        phase of the EDE scalar are modulated by different coupling constants. Positive 
        coupling constants indicate that energy flows from dark matter to dark energy, 
        thereby increasing the energy density fraction of EDE, while negative coupling 
        constants lead to a decrease in the energy density fraction of EDE.}
    \label{fig:2}
\end{figure*}

Fig.~\ref{fig:2} illustrates the evolution of the EDE scalar (left panel) and the fraction 
of the EDE energy density to the total energy density (right panel) with respect to the 
scale factor.
We fix the mass of SFDM at $10^{-22}$ eV and vary the coupling constant, 
while the other cosmological parameters, including the EDE parameter, are given by 
equation~(\ref{eq:ede}). 
Slight variations in the amplitude and phase of the evolution of the EDE scalar field 
are induced by different coupling constants. The sign of the coupling constant 
determines the direction of energy transfer. Positive coupling constants indicate 
energy flow from dark matter to dark energy, resulting in an increase in the energy 
density fraction of EDE, as demonstrated in the right panel of the figure.

\subsection{Perturbution Equations}
We employ the synchronous gauge to compute the perturbation equations for both EDE 
and SFDM. The line element is defined as, 
\begin{equation}
    ds^2=-dt^2+a^2(t)(\delta_{ij}+h_{ij})dx^idx^j.
\end{equation}
The perturbed Klein-Gordon equation in the Fourier mode are given by
\begin{equation}
	\begin{split}
		\ddot{\delta\phi}&=-3H\dot{\delta\phi}-[\frac{k^2}{a^2}+V_{\phi\phi}-\frac{1}{2}(\frac{\lambda}{M_{pl}})^2\exp(\frac{\lambda\phi}{M_{pl}})\dot{\chi}^2]\delta\phi\\
		&-\frac{1}{2}\dot{h}\dot{\phi}+\frac{\lambda}{M_{pl}}\exp(\frac{\lambda\phi}{M_{pl}})\dot{\chi}\dot{\delta\chi},
	\end{split}
\end{equation}

\begin{equation}
	\begin{split}
        \ddot{\delta\chi}&=-(3H+\frac{\lambda\dot{\phi}}{M_{pl}})\dot{\delta\chi}-[\frac{k^2}{a^2}+m_{\chi}^2\exp(-\frac{\lambda\phi}{M_{pl}})]\delta\chi-\frac{1}{2}\dot{h}\dot{\chi}\\
		&+\frac{\lambda}{M_{pl}}[m_{\chi}^2\chi\exp(-\frac{\lambda\phi}{M_{pl}})\delta\phi-\dot{\chi}\dot{\delta\phi}],
	\end{split}
\end{equation}
where $V_{\phi\phi}$ represents the second order partial derivative of the EDE 
potential with respect to $\phi$.

According to \citet{PRD.58.023503,Hu_1998}, the density perturbation, pressure 
perturbation, and velocity divergence of SFDM can be expressed as,
\begin{subequations}
    \begin{align}
        &\delta\rho_{\chi}=\zeta\dot{\chi}\dot{\delta\chi}+m_{\chi}^2\chi\delta\chi,\\
        &\delta p_{\chi}=\zeta\dot{\chi}\dot{\delta\chi}-m_{\chi}^2\chi\delta\chi,\\
        &(\rho_{\chi}+p_{\chi})\Theta_{\chi}=\frac{k^2}{a}\zeta\dot{\chi}\delta\chi.
    \end{align} 
    \label{eqrpv}
\end{subequations}
We employ some new variables to compute the perturbation equations of SFDM \citet{PRD.96.061301},
\begin{subequations}
    \begin{align}
        &\sqrt{\Omega_{\chi}}(\delta_0\sin{\frac{\theta}{2}}+\delta_1\cos{\frac{\theta}{2}})=\sqrt{\frac{2}{3}}\frac{\sqrt{\zeta}\dot{\delta\chi}}{M_{pl}H},\\
        &\sqrt{\Omega_{\chi}}(\delta_1\sin{\frac{\theta}{2}}-\delta_0\cos{\frac{\theta}{2}})=\sqrt{\frac{2}{3}}\frac{m_{\chi}\delta\chi}{M_{pl}H}.
    \end{align} 
\end{subequations}
One can derive the evolution equation for the new variable,
\begin{equation}
    \begin{split}
        \dot{\delta_0}&=-\frac{\lambda Hy_1\delta\phi}{2M_{pl}\sqrt{\zeta}}\sin\theta-(\frac{\lambda\dot{\delta\phi}}{M_{pl}}+\frac{1}{2}\dot{h})(1-\cos\theta)\\
        &+\delta_0H\omega\sin\theta-\delta_1[(3H+\frac{\lambda\dot{\phi}}{2M_{pl}})\sin\theta+H\omega(1-\cos\theta)],
    \end{split}
\end{equation}

\begin{equation}
    \begin{split}
        \dot{\delta_1}&=-\frac{\lambda Hy_1\delta\phi}{2M_{pl}\sqrt{\zeta}}(1+\cos\theta)-(\frac{\lambda\dot{\delta\phi}}{M_{pl}}+\frac{1}{2}\dot{h})\sin\theta\\
        &+\delta_0H\omega(1+\cos\theta)-\delta_1[(3H+\frac{\lambda\dot{\phi}}{2M_{pl}})\cos\theta+H\omega\sin\theta],
    \end{split}
\end{equation} 
where 
\begin{equation}
    \omega=\frac{k^2\sqrt{\zeta}}{2a^2m_{\chi}H}=\frac{k^2\sqrt{\zeta}}{a^2H^2y_1}.
\end{equation}
The relationship between the density perturbations, pressure perturbations, and velocity 
divergence of SFDM and the new variables can be deduced from equation~(\ref{eqrpv}),
\begin{subequations}
    \begin{align}
        &\delta\rho_{\chi}=\rho_{\chi}\delta_0,\\
        &\delta p_{\chi}=\rho_{\chi}(\delta_1\sin\theta-\delta_0\cos\theta),\\
        &(\rho_{\chi}+p_{\chi})\Theta_{\chi}=\frac{k^2\sqrt{\zeta}}{aHy_1}\rho_{\chi}[\delta_1(1-\cos\theta)-\delta_0\sin\theta].
    \end{align}
\end{subequations}

\subsection{Initial Conditions}
In the early universe, Hubble friction in the scalar fields dominated and both 
EDE and SFDM were effectively frozen, undergoing a slow-roll process. The initial 
value of $\dot{\phi}$ can be set to zero, the term in the EDE equation containing 
$\dot{\chi}$ can be neglected during that period. 
Therefore, the equations of motion for EDE and SFDM simplify to an uncoupled form 
(the equation for the new variable $\theta$ is an exception, and we will address 
this point later). 
We introduce the ratio of the initial value of the EDE scalar to the axion decay 
constant, $\alpha_i=\phi_i/f_{\phi}$ as the model parameter \citet{PRD.102.043507,PRD.101.063523}.
We refer to the initial conditions of uncoupled SFDM and make modifications, 
\begin{equation}
    y_{1i}=\frac{2m_{\chi}}{H_0\sqrt{\Omega_r a_{i}^{-4}}},
    \quad
    \theta_{i}=\frac{1}{5}y_{1i}\exp{(\frac{-\lambda\phi_{i}}{2M_{pl}})},
\end{equation}
where $\Omega_r$ represents the energy density fraction of all radiation components 
at present, and $a_i$ denotes the initial value of the scale factor. 
It should be noted that due to coupling, the initial value of $\theta$ is multiplied 
by an additional factor, $\exp{\frac{-\lambda\phi_{i}}{2M_{pl}}}$. For the derivation 
of the initial conditions of the original SFDM model, please refer to \citet{Ure_a_L_pez_2016, PRD.96.061301}. 
Based on the current value of the dark matter energy density, we employ the widely 
used shooting method in the Boltzmann code \texttt{CLASS}\footnote{\url{http://class-code.net}} 
\cite{Blas_2011,1104.2932} to obtain the initial value of $\Omega_{\chi}$. 
For the perturbation equations of EDE and SFDM, we employ adiabatic initial conditions, 
with detailed descriptions provided in \citet{PRD.101.063523} and \citet{PRD.96.061301}. 

\section{Numerical Results}
\label{sec:nr}
Based on the description provided in the previous section, we modified the publicly 
available Boltzmann code \texttt{CLASS} \cite{Blas_2011,1104.2932} to incorporate the new model.

We present numerical results using the cosmological parameters borrowed from Table IV 
in \citet{PRD.106.043525}. More specifically, for the $\Lambda$CDM model, we utilise 
the following parameter values:
\begin{align}
    &\omega_\mathrm{b}=0.02258, \quad \omega_\mathrm{dm}=0.1176,\\
    \notag
    &H_0=68.47, \quad \ln(10^{10}A_\mathrm{s})=3.041,\\
    \notag
    &n_\mathrm{s}=0.9706, \quad \tau_\mathrm{reio}=0.0535.
\end{align}
For the KCS model, we fix the mass of the SFDM at $10^{-22}$ eV, while varying the 
coupling constant $\lambda$ to be 0, -0.03, and 0.03. All other cosmological parameters are 
chosen based on the results of the EDE model,
\begin{align}
    \label{eq:ede}
    &\omega_\mathrm{b}=0.02281, \quad \omega_\mathrm{dm}=0.1287,\\
    \notag
    &H_0=72.02, \quad \ln(10^{10}A_\mathrm{s})=3.065,\\
    \notag
    &n_\mathrm{s}=0.9895, \quad \tau_\mathrm{reio}=0.0581, \quad \alpha_i=2.77,\\
    \notag
    &\log_{10}(f_{\phi})=26.61, \quad \log_{10}(m_{\phi})=-27.31.    
\end{align}

\begin{figure}
	\includegraphics[width=\columnwidth]{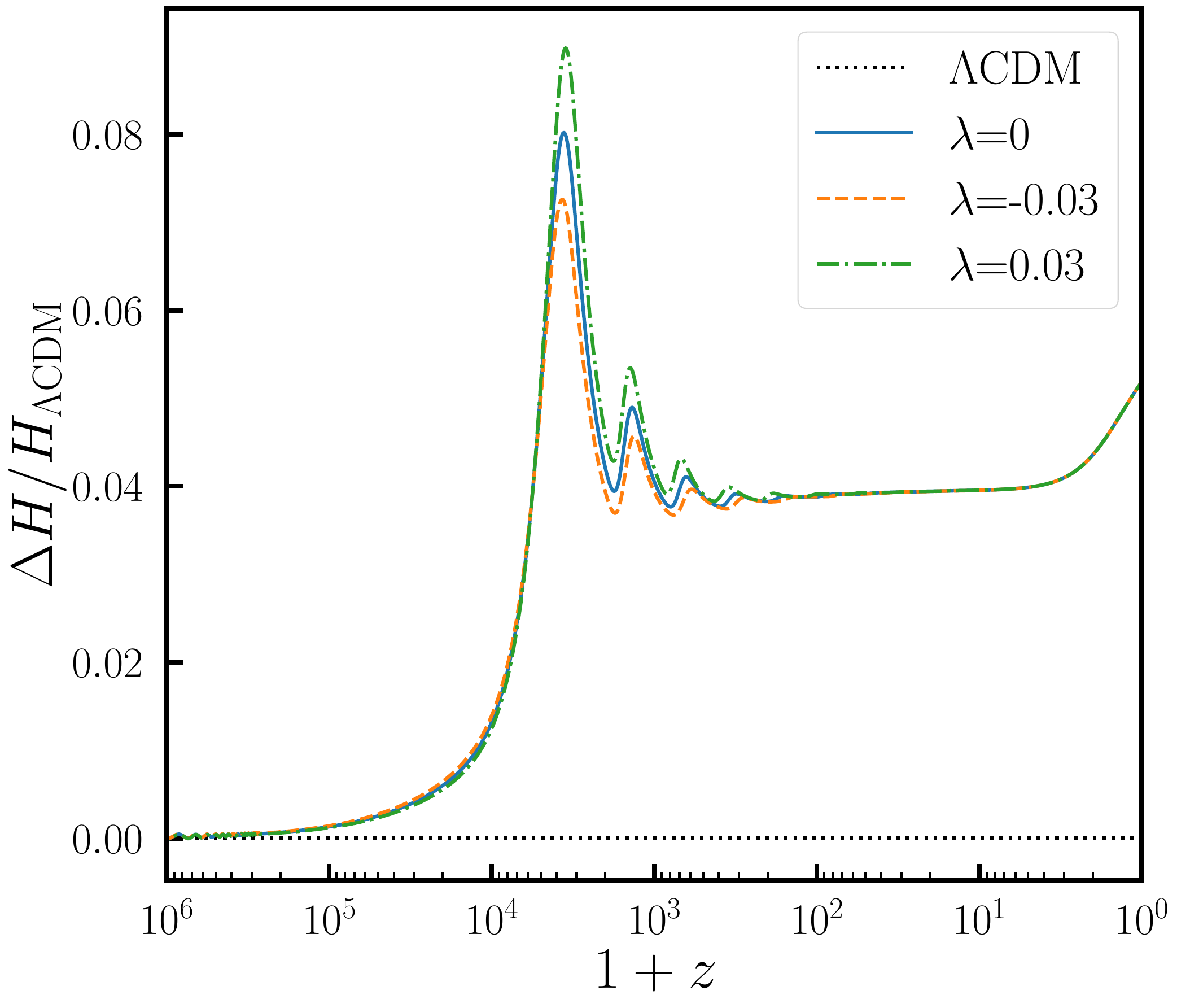}
    \caption{
        The evolution of the difference in the Hubble parameter between 
        the KCS model and the $\Lambda$CDM model with redshift.
        The inclusion of an early dark energy component in the KCS models results 
        in higher values of the Hubble parameter compared to the $\Lambda$CDM model. 
        Furthermore, distinct coupling constants further influence the 
        evolution of the Hubble parameter. Negative coupling constants decrease the Hubble 
        parameter, while positive coupling constants increase it.}
    \label{fig:3}
\end{figure}

We present in Fig.~\ref{fig:3} the evolution of the the differences in the Hubble 
parameter between the KCS model and the $\Lambda$CDM model for various coupling 
constants with redshift.
The $\Lambda$CDM model is depicted by the black dotted line, while the results for the 
KCS model with coupling parameters 0, -0.03, and 0.03 are represented by the blue solid, 
orange dashed, and green dash-dotted lines, respectively.

The presence of an early dark energy component in the KCS models leads to 
higher values of the Hubble parameter compared to the $\Lambda$CDM model. 
Different coupling constants further influence the evolution of the Hubble parameter, 
with positive coupling constants indicating a transfer of energy density from dark 
matter to dark energy, resulting in an increase in the EDE energy density fraction (as 
shown in Fig.~\ref{fig:2}), leading to a larger Hubble parameter near the critical 
redshift $z_\mathrm{c}$. Conversely, negative coupling constants have the opposite 
effect, reducing the Hubble parameter.

\begin{figure}
	\includegraphics[width=\columnwidth]{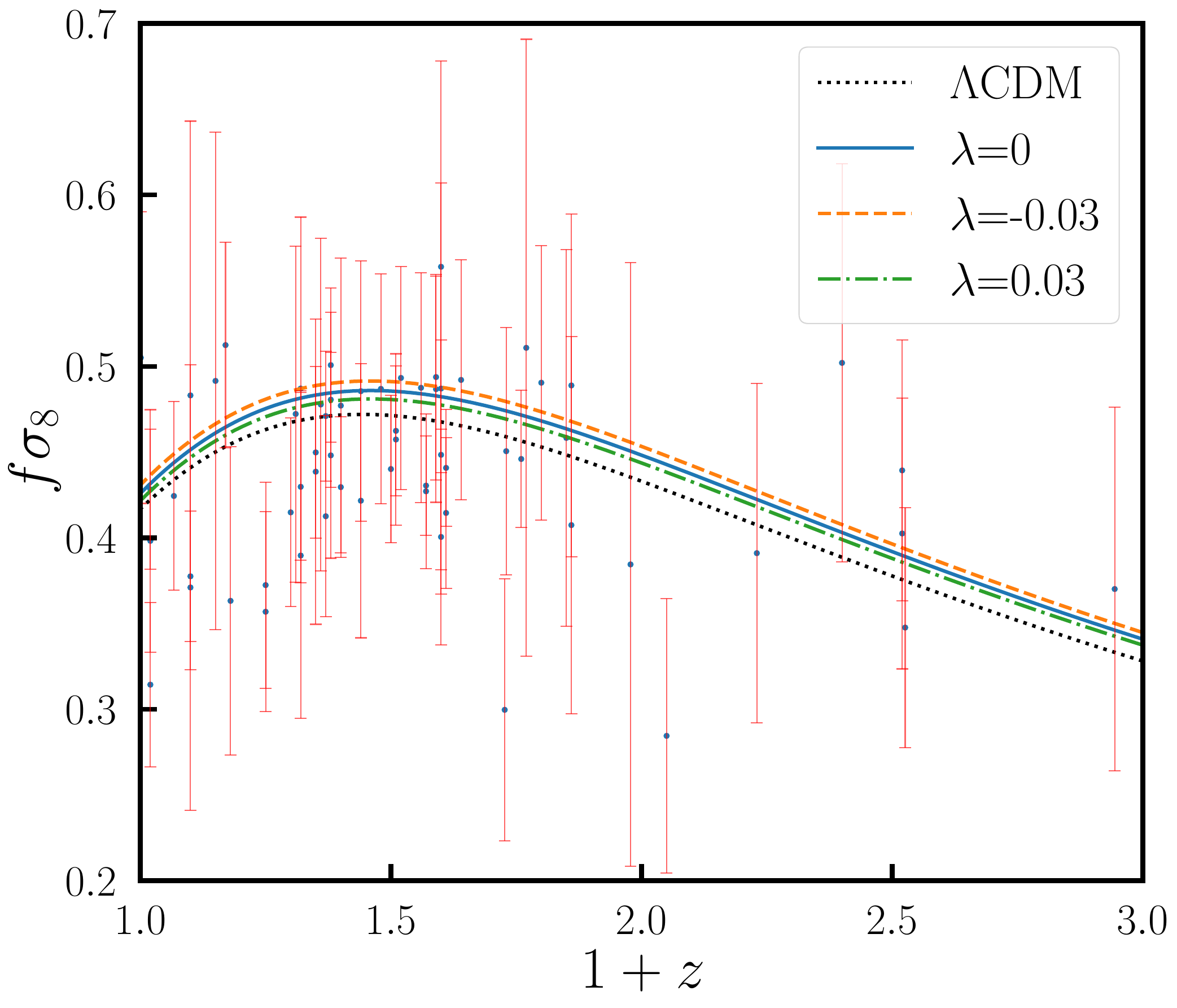}
    \caption{
		The redshift evolution of $f\sigma_8$ for the various models is examined. 
        In the KCS model, the impact of the EDE component leads to a higher value of $f\sigma_8$ 
        compared to the results obtained within the $\Lambda$CDM model. However, when the 
        coupling constant is non-zero (indicating the existence of interaction between dark 
        matter and dark energy), it serves to amend this effect. A positive value of the 
        coupling constant mitigates the adverse effects introduced by EDE.
        }
    \label{fig:4}
\end{figure}

Fig.~\ref{fig:4} illustrates the redshift evolution of $f\sigma_8$. The 63 observed 
Redshift Space Distortion $f\sigma_8(z)$ data points are gathered 
from \citet{PRD.97.103503}.
The KCS models yield higher $f\sigma_8$ compared to the $\Lambda$CDM model, 
exacerbating the existing $S_8$ tension. 
This primarily stems from the effects of the EDE component in the KCS model. However, 
when the coupling constant is non-zero, indicating an interaction between dark matter 
and dark energy, this result can change. In the case of a positive coupling 
constant (green dash-dotted line), the value of $f\sigma_8$ is less than the result 
when the coupling constant is zero (blue solid line), while the effect is opposite when 
the coupling constant is negative (orange dashed line).    

\begin{figure}
	\includegraphics[width=\columnwidth]{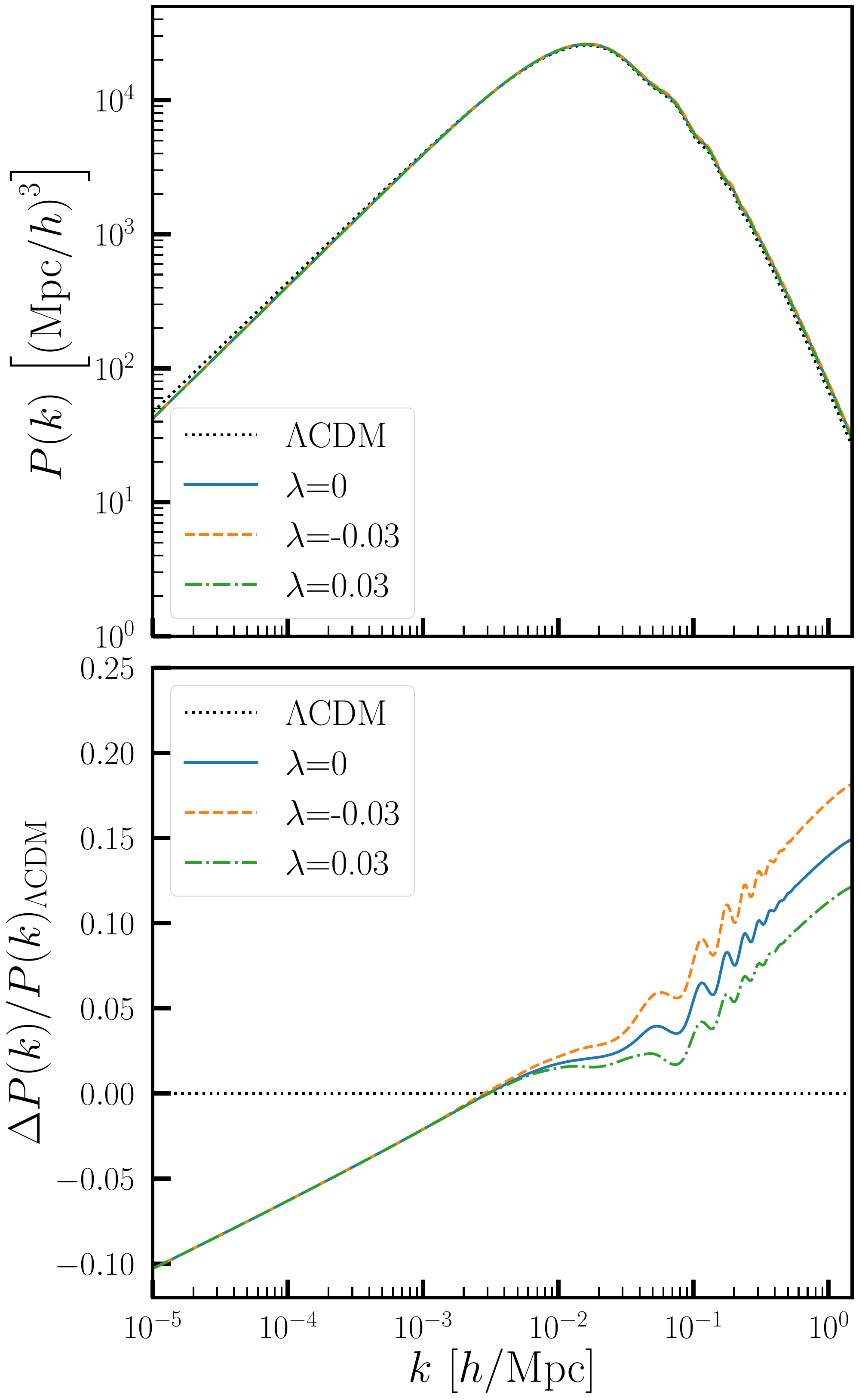}
    \caption{
		The linear matter power spectra of various models (top panel) and their 
        deviations from the $\Lambda$CDM model (bottom panel) are presented. 
        The EDE component in the KCS model increases the matter 
        power on small scales, but the interaction between dark matter and dark 
        energy modifies this result, with positive coupling constants reducing 
        this effect.}
    \label{fig:5}
\end{figure}

In Fig.~\ref{fig:5}, we present the linear 
matter power spectra for various models (top panel) and the differences between the 
power spectra of KCS model relative to the $\Lambda$CDM model (bottom panel).

Similar to previous discussions, the EDE component in the KCS model amplifies the 
matter power spectrum on small scales, exacerbating the large-scale structure tension. 
However, the interaction between dark matter and dark energy can rectify this outcome. 
A positive coupling constant leads to the flow of dark matter energy density towards 
dark energy, suppressing the growth of matter structure and resulting in a smaller 
power spectrum on small scales.

\section{Data and Methodology}
\label{sec:dm}
We conducted the Markov Chain Monte Carlo (MCMC) analysis using \texttt{MontePython}
\footnote{\url{https://github.com/baudren/montepython_public}} \citet{Audren_2013,brinck} 
to obtain the posterior distribution of the model 
parameters.
The MCMC chains were analysed using \texttt{GetDist}
\footnote{\url{https://github.com/cmbant/getdist}} \citet{lewis2019getdist}.

\subsection{Datasets}
To perform the MCMC analysis, we used the following datasets: 
\begin{itemize}
    \item[1.] \textbf{CMB}: The temperature and polarization power spectra obtained from 
    the \textit{Planck} 2018 low-$\ell$ and high-$\ell$ measurements, as well as the CMB 
    lensing power spectrum 
    \citet{osti_1676388,osti_1775409,planck2020}.
    \item[2.] \textbf{BAO}: The measurements obtained from the BOSS-DR12 $f\sigma_8$ 
    sample comprise the combined LOWZ and CMASS galaxy samples \citet{Alam_2017,Buen_Abad_2018}, 
    as well as the small-z measurements derived from 6dFGS and the SDSS DR7 \citet{19250.x,stv154}.
    \item[3.] \textbf{Supernovae}: The Pantheon dataset consists of 1048 supernovae type 
    Ia with redshift values ranging from 0.01 to 2.3 \citet{Scolnic_2018}.
\end{itemize}  

By combining CMB and BAO data, acoustic horizon measurements can be made at multiple 
redshifts, breaking geometric degeneracies and constraining the physical processes 
between recombination and the redshift at which BAO is measured. The supernova data 
obtained from the Pantheon sample significantly constrains late-time new physic within 
its measured redshift range. 

\begin{itemize}
    \item[4.] \textbf{SH0ES}: According to the latest SH0ES measurement, the value of the 
    Hubble constant is estimated to be $73.04 \pm 1.04$\,km\,/\,s\,/\,Mpc \citet{Riess_2022}.
    \item[5.] \textbf{DES-Y3}: The Dark Energy Survey Year-3 has yielded valuable data on 
    weak lensing and galaxy clustering, from which the parameter $S_8$ has been measured 
    to be $0.776 \pm 0.017$ \citet{PRD.105.023520}.
\end{itemize}

We employ the $H_0$ measurements from SH0ES to 
alleviate the prior volume effect \citet{PRD.103.123542} and 
assess the ability of the novel model to mitigate the tension between $H_0$ local 
measurement and CMB inference result. Additionally, we incorporate the $S_8$ data from 
DES-Y3 to investigate the efficacy of the model in easing the large-scale structure tension.

\subsection{Results}
The results of parameter constraints are presented in Table~\ref{tab1}, where we utilised 
a comprehensive dataset comprising CMB, BAO, SNIa, SH0ES, and $S_8$ from DES-Y3 data to 
individually constrain the $\Lambda$CDM, EDE, and KCS models. 
\begin{table*}
    \centering
    \caption{The table illustrates the best-fit parameters and 68\% confidence level 
    marginalised constraints for the $\Lambda$CDM, EDE, and KCS models. These constraints 
    are derived from comprehensive data sets including CMB, BAO, SNIa, SH0ES, and $S_8$ 
    measurements obtained from DES-Y3. The upper portion of the table consists of the 
    cosmological parameters that were explored using MCMC sampling, while the lower 
    portion presents the derived parameters. }
    \label{tab1}
    \renewcommand{\arraystretch}{1.2}
\begin{tabular} { l  c  c  c}
    \hline
    \hline
    Model  &  $\Lambda$CDM  &  EDE  &  KCS\\
    \hline
    $100\omega{}_\mathrm{b }$&
    $2.260(2.263\pm 0.014)$&
    $2.276(2.281^{+0.024}_{-0.020})$&
    $2.305(2.281\pm 0.020)$\\

    $\omega{}_\mathrm{dm}$&
    $0.11729(0.11725\pm 0.00084)$&
    $0.1310(0.1299\pm 0.0028)$&
    $0.1286(0.1303\pm 0.0028)$\\

    $H_0$&
    $68.64(68.71^{+0.35}_{-0.41})$&
    $71.85(72.46\pm 0.86)$&
    $72.71(72.38^{+0.71}_{-0.82})$\\

    $\ln(10^{10}A_\mathrm{s})$&
    $3.047(3.050\pm 0.015)$&
    $3.057(3.063^{+0.015}_{-0.017})$&
    $3.069(3.057\pm 0.018)$\\

    $n_\mathrm{s}$&
    $0.9733(0.9722\pm 0.0040)$&
    $0.9877(0.9908\pm 0.0059)$&
    $0.9943(0.9875^{+0.0065}_{-0.0041})$\\

    $\tau{}_\mathrm{reio}$&
    $0.0575(0.0592\pm 0.0082)$&
    $0.0539(0.0563\pm 0.0090)$&
    $0.0614(0.0543\pm 0.0093)$\\

    $\log_{10}(m_{\phi})$&
    $-$&
    $-27.292(-27.290\pm 0.055)$&
    $-27.241(-27.295^{+0.052}_{-0.042})$\\

    $\log_{10}(f_{\phi})$&
    $-$&
    $26.632(26.616^{+0.056}_{-0.033})$&
    $26.632(26.645\pm 0.051)$\\

    $\alpha_\mathrm{i} $&
    $-$&
    $2.762(2.783\pm 0.069)$&
    $2.734(2.696^{+0.069}_{-0.055})$\\    

    $\lambda$&
    $-$&
    $-$&
    $0.0112(0.030\pm 0.026)$\\

    $\log_{10}(m_{\chi})$&
    $-$&
    $-$&
    $-21.980(-21.988\pm 0.063)$\\  
    
    \hline

    $10^{-9}A_\mathrm{s}$&
    $2.105(2.112\pm 0.032)$&
    $2.127(2.139^{+0.031}_{-0.036})$&
    $2.153(2.127\pm 0.038)$\\

    $100\theta{}_\mathrm{s}$&
    $1.04206(1.04217^{+0.00025}_{-0.00031})$&
    $1.04121(1.04145\pm0.00043)$&
    $1.04149(1.04140^{+0.00028}_{-0.00039})$\\

    $f_\mathrm{EDE} $&
    $-$&
    $0.1183(0.119^{+0.023}_{-0.018})$&
    $0.1168(0.124\pm 0.021)$\\

    $\log_{10}(z_\mathrm{c})$&
    $-$&
    $3.571(3.568\pm 0.034)$&
    $3.601(3.569^{+0.028}_{-0.025})$\\    

    $\Omega{}_\mathrm{m}$&
    $0.2983(0.2977\pm 0.0048)$&
    $0.2991(0.2923\pm 0.0056)$&
    $0.2880(0.2935^{+0.0057}_{-0.0043})$\\

    $\sigma_8$&
    $0.8039(0.8047\pm 0.0060)$&
    $0.8329(0.8325\pm 0.0083)$&
    $0.8313(0.8284^{+0.0088}_{-0.0071})$\\    

    $S_{8}$&
    $0.8016(0.8016^{+0.0096}_{-0.0080})$&
    $0.8316(0.822^{+0.011}_{-0.0093})$&
    $0.8146(0.819^{+0.013}_{-0.0092})$\\

    \hline      
    $\chi^2_\mathrm{tot}$  &  $3838.20$  &  $3826.46$  &  $3825.42$\\      
    $\Delta \mathrm{AIC}$        &  $-$  &  $ -5.74$  &  $ -2.78$\\ 
    \hline
    \hline
\end{tabular}
\end{table*}
The upper segment of the table represents the parameters subjected to MCMC sampling, 
while the lower segment displays the derived parameters.

Firstly, it is noteworthy that the KCS model constrains the coupling constant $\lambda$ 
to be $0.030 \pm 0.026$ at a 68\% confidence level, with a best-fit value of 0.0112. 
This indicates an interaction between dark matter and dark energy, specifically the 
conversion of dark matter to dark energy.

From the perspective of the Hubble constant, the EDE model and KCS model yield $H_0$ 
values of $72.46 \pm 0.86$\,km\,/\,s\,/\,Mpc and $72.38^{+0.71}_{-0.82}$\,km\,/\,s\,/\,Mpc, 
respectively, at a 68\% confidence level, both of which exceed the value of 
$68.71^{+0.35}_{-0.41}$\,km\,/\,s\,/\,Mpc obtained by the $\Lambda$CDM model. Therefore, 
both the EDE model and KCS model demonstrate the capacity to address the Hubble tension.

However, the performance of both models on $S_8$ is suboptimal. The best-fit values of 
$S_8$ for the EDE and KCS models are 0.8316 and 0.8146, respectively, whereas $\Lambda$CDM 
yields a result of0.8016. Both models exacerbate the preexisting $S_8$ tension. 
The result obtained from the KCS model is smaller than that from the EDE model, 
indicating that the new model partially alleviates the negative effect of the EDE model. However, 
when comparing the marginalised posterior probabilities, this feature vanishes, both models yield 
similar results. The constrained values of $S_8$ obtained from the EDE model and the KCS model are 
$0.822^{+0.011}_{-0.0093}$ and $0.819^{+0.013}_{-0.0092}$, respectively, at a 68\% confidence level.

These discussions can be visually depicted in the marginalised posterior distributions of 
various models, as illustrated in Fig.~\ref{fig:6}. 
The complete posterior distributions 
can be found in Fig.~\ref{fig:8} in the Appendix section.
\begin{figure}
	\includegraphics[width=\columnwidth]{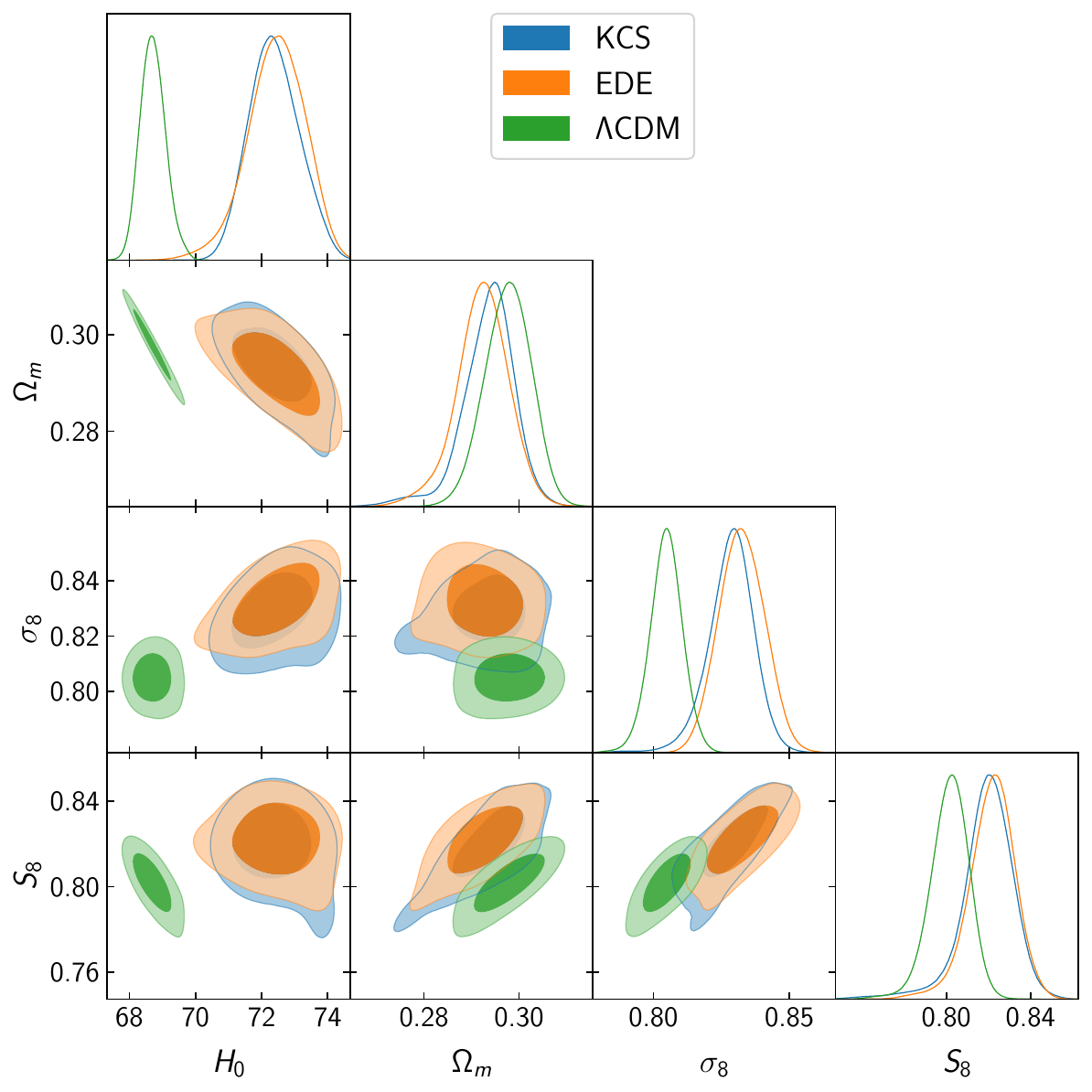}
    \caption{
		The marginalised posterior distributions of the three models are presented. Both 
        the EDE and KCS models yield larger values of $H_0$ and hence, a larger $S_8$ 
        compared to the $\Lambda$CDM model. However, the KCS model partially alleviates 
        the $S_8$ tension when compared to the EDE model.}
    \label{fig:6}
\end{figure}

The penultimate row of Table~\ref{tab1} displays the $\chi^2_\mathrm{tot}$ 
values of the different models. 
The discrepancies of the EDE and KCS models compared to the $\Lambda$CDM model are -11.74 and 
-12.78, respectively. Both models exhibit significantly reduced $\chi^2_\mathrm{tot}$ 
values compared to the $\Lambda$CDM model, primarily due to the contribution from the 
SH0ES data. Furthermore, the KCS model demonstrates a lower $\chi^2_\mathrm{tot}$ value 
than the EDE model, attributed to its smaller $S_8$ value, which better aligns with the 
DES-Y3 data.

We also calculated the Akaike Information Criterion (AIC) for model comparison \citet{1100705}, 
\begin{equation}
    \mathrm{AIC}=\chi^2_\mathrm{tot}+2k,
\end{equation}
where $k$ represents the number of fitting parameters. The results of different models 
compared to the $\Lambda$CDM model are presented in the 
last row of Table~\ref{tab1}. We find that the EDE model has the lowest AIC, followed 
by the KCS model, both of which yield smaller values than that of the $\Lambda$CDM 
model. This indicates that both models outperform the $\Lambda$CDM model. Furthermore, 
compared to the EDE model, although the KCS model has a smaller $\chi^2_\mathrm{tot}$ 
value, its introduction of additional parameters makes its performance inferior to 
that of the EDE model from the perspective of AIC.

Additionally, we conducted a renewed MCMC analysis after excluding the SH0ES data to 
investigate its impact on the EDE parameters. Our constrained parameter results are 
illustrated in Table~\ref{tab2}, where the utilised data only consist of CMB, BAO, SNIa, 
and $S_8$ measurements obtained from DES-Y3. 
\begin{table*}
    \centering
    \caption{
        After excluding SH0ES data, the best-fit values and 68\% confidence 
    level marginalised constraints on the parameters of the $\Lambda$CDM model, 
    the EDE model, and the KCS model, using only CMB, BAO, SNIa, and $S_8$ 
    measurements obtained from DES-Y3 data.}
    \label{tab2}
    \renewcommand{\arraystretch}{1.2}
\begin{tabular} { l  c  c  c}
    \hline
    \hline
    Model  &  $\Lambda$CDM  &  EDE  &  KCS\\
    \hline
    $100\omega{}_\mathrm{b }$&
    $2.249(2.252\pm 0.014)$&
    $2.283(2.279\pm 0.018)$&
    $2.257(2.275\pm 0.020)$\\

    $\omega{}_\mathrm{dm}$&
    $0.11823(0.11825\pm 0.00085)$&
    $0.1232(0.1230^{+0.0020}_{-0.0023})$&
    $0.1210(0.1227^{+0.0023}_{-0.0021})$\\

    $H_0$&
    $68.11(68.21\pm 0.39)$&
    $69.74(69.94^{+0.79}_{-0.49})$&
    $68.94(69.85^{+0.73}_{-0.62})$\\

    $\ln(10^{10}A_\mathrm{s})$&
    $3.045(3.045\pm 0.017)$&
    $3.060(3.055^{+0.015}_{-0.019})$&
    $3.044(3.050\pm 0.016)$\\

    $n_\mathrm{s}$&
    $0.9699(0.9693\pm 0.0038)$&
    $0.9845(0.9813^{+0.0056}_{-0.0048})$&
    $0.9714(0.9798\pm 0.0053)$\\

    $\tau{}_\mathrm{reio}$&
    $0.0568(0.0563\pm 0.0080)$&
    $0.0594(0.0577^{+0.0081}_{-0.010})$&
    $0.0531(0.0550\pm 0.0083)$\\

    $\log_{10}(m_{\phi})$&
    $-$&
    $-26.837(-26.90^{+0.10}_{-0.081})$&
    $-26.992(-26.909\pm 0.072)$\\

    $\log_{10}(f_{\phi})$&
    $-$&
    $26.392(26.405^{+0.081}_{-0.056})$&
    $26.261(26.379^{+0.084}_{-0.048})$\\

    $\alpha_\mathrm{i} $&
    $-$&
    $2.845(2.81^{+0.10}_{-0.076})$&
    $2.779(2.765^{+0.090}_{-0.066})$\\    

    $\lambda$&
    $-$&
    $-$&
    $0.060(0.032^{+0.039}_{-0.028})$\\

    $\log_{10}(m_{\chi})$&
    $-$&
    $-$&
    $-22.608(-22.670\pm 0.059)$\\  
    
    \hline

    $10^{-9}A_\mathrm{s}$&
    $2.101(2.102\pm 0.035)$&
    $2.132(2.122^{+0.032}_{-0.041})$&
    $2.098(2.111\pm 0.033)$\\

    $100\theta{}_\mathrm{s}$&
    $1.04181(1.04204^{+0.00027}_{-0.00030})$&
    $1.04182(1.04183^{+0.00026}_{-0.00041})$&
    $1.04192(1.04195^{+0.00022}_{-0.00046})$\\

    $f_\mathrm{EDE} $&
    $-$&
    $0.0548(0.056\pm 0.018)$&
    $0.0311(0.054^{+0.021}_{-0.015})$\\

    $\log_{10}(z_\mathrm{c})$&
    $-$&
    $3.810(3.780^{+0.050}_{-0.037})$&
    $3.725(3.772\pm 0.038)$\\    

    $\Omega{}_\mathrm{m}$&
    $0.3047(0.3040\pm 0.0050)$&
    $0.3016(0.2995^{+0.0058}_{-0.0047})$&
    $0.3034(0.2996^{+0.0060}_{-0.0049})$\\

    $\sigma_8$&
    $0.8055(0.8056\pm 0.0065)$&
    $0.8216(0.8178\pm 0.0079)$&
    $0.8116(0.8168^{+0.0064}_{-0.0072})$\\    

    $S_{8}$&
    $0.8118(0.8110\pm 0.0093)$&
    $0.8237(0.817\pm 0.010)$&
    $0.8161(0.816^{+0.011}_{-0.0089})$\\

    \hline
    $\chi^2_\mathrm{tot}$  &  $3818.96$  &  $3822.28$  &  $3822.18$\\      
    \hline
    \hline
\end{tabular}
\end{table*}

We observed that in the absence of SH0ES data, the phenomenological parameters $f_\mathrm{EDE}$ 
obtained from the EDE model and KCS model constraints are very small, with best-fit values of 
0.0548 and 0.0311, respectively, which are consistent with previous findings \citet{PRD.103.123542,poulin2023ups}. 
It is noteworthy that in our MCMC analysis, we sampled over the axion mass $m_{\phi}$ and decay 
constant $f_{\phi}$, while the EDE parameters $f_\mathrm{EDE}$ and $z_\mathrm{c}$ are derived 
parameters. As a result, our findings may differ from those of other studies.

\begin{figure}
	\includegraphics[width=\columnwidth]{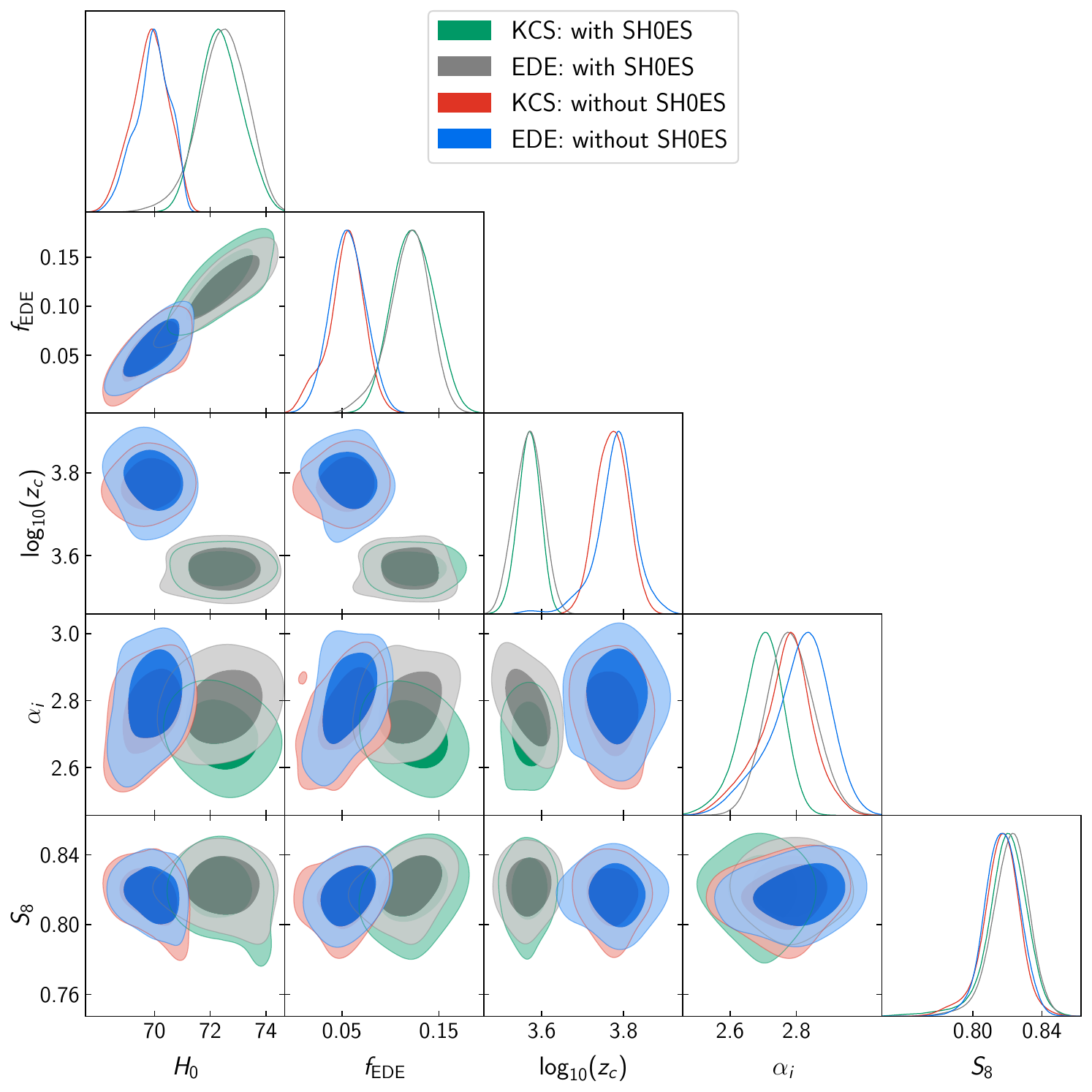}
    \caption{
		The marginalised posterior distributions of the parameters $H_0$, $S_8$, and 
        EDE parameters for the KCS and EDE models are derived with and without the 
        inclusion of SH0ES data. After the exclusion of SH0ES data, both models yield 
        very small values for $f_\mathrm{EDE}$.}
    \label{fig:7}
\end{figure}

In Fig.~\ref{fig:7}, we present the contour plots of the constraints on the $H_0$, $S_8$, and 
EDE parameters for the KCS model and EDE model, with and without the inclusion of SH0ES data.

The impact of the SH0ES data on the EDE parameter is clearly evident. Excluding the 
SH0ES data leads to a reduction in the EDE parameter $f_\mathrm{EDE}$ and an associated 
higher critical redshift $z_\mathrm{c}$, indicating a subtle signal of the existence 
of EDE. The results obtained from the KCS model and the EDE model are consistent.

In order to quantify the level of tension with the SH0ES data, we computed 
the following tension metric (in units of Gaussian $\sigma$) \citet{PhysRevD.99.043506,SCHONEBERG20221}, 
\begin{equation}
    Q_\mathrm{DMAP}\equiv \sqrt{\chi^2(w/\,\mathrm{SH0ES})-\chi^2(w/o\, \mathrm{SH0ES})},
\end{equation}
which calculates the difference in $\chi^2$ when considering with and without SH0ES data. 
This metric effectively captures the non-Gaussianity of the posterior. 
The tension metric yields results of 4.4$\sigma$, 2.1$\sigma$, and 1.8$\sigma$ for 
the $\Lambda$CDM model, EDE model, and KCS model, respectively (it is noted that our 
data includes $S_8$ from DES-Y3, hence yielding slightly different results from 
previous studies). According to this criterion, we consider both the EDE model and 
the KCS model to be superior to the $\Lambda$CDM model.

\section{Conclusions}
\label{sec:con}
In this paper, we investigate the interaction between early dark energy (EDE) and 
scalar field dark matter (SFDM), proposing a kinetically coupled scalar fields (KCS) 
model to alleviate cosmological tensions. The EDE model offers a resolution to the 
Hubble tension, but exacerbates the $S_8$ tension.

In light of this, we propose an interaction between dark matter and dark energy, aiming 
to alleviate the $S_8$ tension through the drag exerted by dark energy on dark matter. 

In particular, motivated by the ability of SFDM to suppress structure growth on small 
scales, we replace cold dark matter with SFDM. Inspired by string theory, we consider 
a kinetic coupling between two scalar fields, where the kinetic term of SFDM is 
multiplied by an EDE scalar-dependent function.

We derived the evolution equations of the KCS model at both the background and 
perturbation levels, and investigated its impact on the evolution of the Hubble 
parameter, as well as on the structure growth and matter power spectrum. By using 
cosmological data from various sources, including the CMB, BAO, SNIa, the SH0ES 
measurement of the Hubble constant, and the DES-Y3 observations, we conducted the MCMC 
analysis.

We obtained a non-zero coupling constant $\lambda$ of $0.030 \pm 0.026$ at a 68\% 
confidence level. This suggests the interaction between dark matter and dark energy, 
with energy transferring from dark matter to dark energy. The values of $H_0$ of the 
EDE model and KCS model are $72.46 \pm 0.86$\,km\,/\,s\,/\,Mpc and 
$72.38^{+0.71}_{-0.82}$\,km\,/\,s\,/\,Mpc, respectively, while the result of 
$\Lambda$CDM model is $68.71^{+0.35}_{-0.41}$\,km\,/\,s\,/\,Mpc. Both models alleviate 
the Hubble tension. 

However, the corresponding cost is that both the EDE model and KCS model yield larger 
values of $S_8$, with their best-fit values being 0.8316 and 0.8146, respectively, 
which are greater than the result of the $\Lambda$CDM model, 0.8016. 
Compared to the EDE model, the result from the KCS model is smaller, 
indicating that the new model partially mitigates the negative effect of the original 
EDE model. However, this feature disappears when comparing the marginalised posterior 
probabilities, as both models yield similar results. The constrained results for $S_8$ 
from the EDE model and the KCS model are $0.822^{+0.011}_{-0.0093}$ and $0.819^{+0.013}_{-0.0092}$, 
respectively, at a 68\% confidence level.

We computed the $\Delta\chi^2_\mathrm{tot}$ values of the EDE model and KCS model relative to 
the $\Lambda$CDM model, which are -11.74 and -12.78, respectively. This is primarily 
attributed to the contribution of SH0ES data. The $\chi^2_\mathrm{tot}$ value of the 
KCS model is slightly smaller than that of the EDE model, mainly due to the fact that 
the KCS model yields a smaller $S_8$, which aligns better with the DES-Y3 data.   
We further calculated the AIC to compare the models, and the results indicate that the 
EDE model has the lowest AIC, followed by the KCS model. The introduction of additional 
parameters in the KCS model leads to its inferior performance compared to the EDE model. 

We also briefly investigated the impact of the SH0ES data on the EDE parameter. When 
excluding the SH0ES data, the constrained $f_\mathrm{EDE}$ parameters from both the 
EDE and KCS models are smaller, suggesting a weak signal of the presence of EDE, 
consistent with previous findings in the literature.

Building upon the EDE model, we considered the coupling between dark matter and 
dark energy to mitigate the negative effect inherent in the EDE model. However, 
the coupled model still falls short of resolving all cosmological tensions, 
and it is not fully supported by the data.
Further efforts are still required to address these challenges comprehensively.

\section*{Acknowledgements}

This work is supported in part by National Natural Science Foundation of China
under Grant No.12075042, Grant No.11675032 (People's Republic of China).

\section*{Data Availability}
The data supporting the findings of this article are available upon reasonable 
request to the corresponding author.
 



\bibliographystyle{mnras}
\bibliography{zkcs} 

\begin{thebibliography}{}
\makeatletter
\relax
\def\mn@urlcharsother{\let\do\@makeother \do\$\do\&\do\#\do\^\do\_\do\%\do\~}
\def\mn@doi{\begingroup\mn@urlcharsother \@ifnextchar [ {\mn@doi@} {\mn@doi@[]}}
\def\mn@doi@[#1]#2{\def\@tempa{#1}\ifx\@tempa\@empty \href {http://dx.doi.org/#2} {doi:#2}\else \href {http://dx.doi.org/#2} {#1}\fi \endgroup}
\def\mn@eprint#1#2{\mn@eprint@#1:#2::\@nil}
\def\mn@eprint@arXiv#1{\href {http://arxiv.org/abs/#1} {{\tt arXiv:#1}}}
\def\mn@eprint@dblp#1{\href {http://dblp.uni-trier.de/rec/bibtex/#1.xml} {dblp:#1}}
\def\mn@eprint@#1:#2:#3:#4\@nil{\def\@tempa {#1}\def\@tempb {#2}\def\@tempc {#3}\ifx \@tempc \@empty \let \@tempc \@tempb \let \@tempb \@tempa \fi \ifx \@tempb \@empty \def\@tempb {arXiv}\fi \@ifundefined {mn@eprint@\@tempb}{\@tempb:\@tempc}{\expandafter \expandafter \csname mn@eprint@\@tempb\endcsname \expandafter{\@tempc}}}

\bibitem[\protect\citeauthoryear{Abbott, Aguena, Alarcon  et~al.}{Abbott et~al.}{2022}]{PRD.105.023520}
Abbott T. M.~C.,  Aguena M.,  Alarcon A.,   et~al., 2022, \mn@doi [Phys. Rev. D] {10.1103/PhysRevD.105.023520}, 105, 023520

\bibitem[\protect\citeauthoryear{Aghanim, Akrami, Ashdown  et~al.}{Aghanim et~al.}{2020b}]{osti_1775409}
Aghanim N.,  Akrami Y.,  Ashdown M.,   et~al., 2020b, \mn@doi [Astronomy and Astrophysics] {10.1051/0004-6361/201833886}, 641

\bibitem[\protect\citeauthoryear{Aghanim, Akrami, Ashdown  et~al.}{Aghanim et~al.}{2020a}]{osti_1676388}
Aghanim N.,  Akrami Y.,  Ashdown M.,   et~al., 2020a, \mn@doi [Astronomy and Astrophysics] {10.1051/0004-6361/201936386}, 641

\bibitem[\protect\citeauthoryear{Akaike}{Akaike}{1974}]{1100705}
Akaike H.,  1974, \mn@doi [IEEE Transactions on Automatic Control] {10.1109/TAC.1974.1100705}, 19, 716

\bibitem[\protect\citeauthoryear{Alam, Ata, Bailey  et~al.}{Alam et~al.}{2017}]{Alam_2017}
Alam S.,  Ata M.,  Bailey S.,   et~al., 2017, \mn@doi [Monthly Notices of the Royal Astronomical Society] {10.1093/mnras/stx721}, 470, 2617

\bibitem[\protect\citeauthoryear{Alestas \& Perivolaropoulos}{Alestas \& Perivolaropoulos}{2021}]{stab1070}
Alestas G.,  Perivolaropoulos L.,  2021, \mn@doi [Monthly Notices of the Royal Astronomical Society] {10.1093/mnras/stab1070}, 504, 3956

\bibitem[\protect\citeauthoryear{Alexander \& McDonough}{Alexander \& McDonough}{2019}]{ALEXANDER2019134830}
Alexander S.,  McDonough E.,  2019, \mn@doi [Physics Letters B] {https://doi.org/10.1016/j.physletb.2019.134830}, 797, 134830

\bibitem[\protect\citeauthoryear{Alexander, Bernardo  \& Toomey}{Alexander et~al.}{2023}]{Alexander_2023}
Alexander S.,  Bernardo H.,   Toomey M.~W.,  2023, \mn@doi [Journal of Cosmology and Astroparticle Physics] {10.1088/1475-7516/2023/03/037}, 2023, 037

\bibitem[\protect\citeauthoryear{Audren, Lesgourgues, Benabed  \& Prunet}{Audren et~al.}{2013}]{Audren_2013}
Audren B.,  Lesgourgues J.,  Benabed K.,   Prunet S.,  2013, \mn@doi [Journal of Cosmology and Astroparticle Physics] {10.1088/1475-7516/2013/02/001}, 2013, 001

\bibitem[\protect\citeauthoryear{Benevento, Hu  \& Raveri}{Benevento et~al.}{2020}]{PRD.101.103517}
Benevento G.,  Hu W.,   Raveri M.,  2020, \mn@doi [Phys. Rev. D] {10.1103/PhysRevD.101.103517}, 101, 103517

\bibitem[\protect\citeauthoryear{Berghaus \& Karwal}{Berghaus \& Karwal}{2020}]{PRD.101.083537}
Berghaus K.~V.,  Karwal T.,  2020, \mn@doi [Phys. Rev. D] {10.1103/PhysRevD.101.083537}, 101, 083537

\bibitem[\protect\citeauthoryear{Beutler, Blake, Colless  et~al.}{Beutler et~al.}{2011}]{19250.x}
Beutler F.,  Blake C.,  Colless M.,   et~al., 2011, \mn@doi [Monthly Notices of the Royal Astronomical Society] {10.1111/j.1365-2966.2011.19250.x}, 416, 3017

\bibitem[\protect\citeauthoryear{Blanchard \& Ili{\'{c}}}{Blanchard \& Ili{\'{c}}}{2021}]{Blanchard_2021}
Blanchard A.,  Ili{\'{c}} S.,  2021, \mn@doi [Astronomy and Astrophysics] {10.1051/0004-6361/202140974}, 656, A75

\bibitem[\protect\citeauthoryear{Blas, Lesgourgues  \& Tram}{Blas et~al.}{2011}]{Blas_2011}
Blas D.,  Lesgourgues J.,   Tram T.,  2011, \mn@doi [Journal of Cosmology and Astroparticle Physics] {10.1088/1475-7516/2011/07/034}, 2011, 034

\bibitem[\protect\citeauthoryear{Brinckmann \& Lesgourgues}{Brinckmann \& Lesgourgues}{2018}]{brinck}
Brinckmann T.,  Lesgourgues J.,  2018, MontePython 3: boosted MCMC sampler and other features (\mn@eprint {arXiv} {1804.07261})

\bibitem[\protect\citeauthoryear{Buen-Abad, Schmaltz, Lesgourgues  \& Brinckmann}{Buen-Abad et~al.}{2018}]{Buen_Abad_2018}
Buen-Abad M.~A.,  Schmaltz M.,  Lesgourgues J.,   Brinckmann T.,  2018, \mn@doi [Journal of Cosmology and Astroparticle Physics] {10.1088/1475-7516/2018/01/008}, 2018, 008

\bibitem[\protect\citeauthoryear{Cede\~no, Gonz\'alez-Morales  \& Ure\~na L\'opez}{Cede\~no et~al.}{2017}]{PRD.96.061301}
Cede\~no F. X.~L.,  Gonz\'alez-Morales A.~X.,   Ure\~na L\'opez L.~A.,  2017, \mn@doi [Phys. Rev. D] {10.1103/PhysRevD.96.061301}, 96, 061301

\bibitem[\protect\citeauthoryear{Copeland, Liddle  \& Wands}{Copeland et~al.}{1998}]{PRD.57.4686}
Copeland E.~J.,  Liddle A.~R.,   Wands D.,  1998, \mn@doi [Phys. Rev. D] {10.1103/PhysRevD.57.4686}, 57, 4686

\bibitem[\protect\citeauthoryear{Ferreira}{Ferreira}{2021}]{Ferreira_2021}
Ferreira E. G.~M.,  2021, \mn@doi [The Astronomy and Astrophysics Review] {10.1007/s00159-021-00135-6}, 29

\bibitem[\protect\citeauthoryear{Ferreira \& Joyce}{Ferreira \& Joyce}{1998}]{PRD.58.023503}
Ferreira P.~G.,  Joyce M.,  1998, \mn@doi [Phys. Rev. D] {10.1103/PhysRevD.58.023503}, 58, 023503

\bibitem[\protect\citeauthoryear{Freedman \& Madore}{Freedman \& Madore}{2023}]{Freedman:2023zdo}
Freedman W.~L.,  Madore B.~F.,  2023, {The Cepheid Extragalactic Distance Scale: Past, Present and Future} (\mn@eprint {arXiv} {2308.02474})

\bibitem[\protect\citeauthoryear{Ghosh, Khatri  \& Roy}{Ghosh et~al.}{2020}]{PRD.102.123544}
Ghosh S.,  Khatri R.,   Roy T.~S.,  2020, \mn@doi [Phys. Rev. D] {10.1103/PhysRevD.102.123544}, 102, 123544

\bibitem[\protect\citeauthoryear{Guo, Zhang  \& Zhang}{Guo et~al.}{2019}]{Guo_2019}
Guo R.-Y.,  Zhang J.-F.,   Zhang X.,  2019, \mn@doi [Journal of Cosmology and Astroparticle Physics] {10.1088/1475-7516/2019/02/054}, 2019, 054

\bibitem[\protect\citeauthoryear{Hildebrandt, Köhlinger, Busch  et~al.}{Hildebrandt et~al.}{2020}]{Hildebrandt_2020}
Hildebrandt H.,  Köhlinger F.,  Busch J.,   et~al., 2020, \mn@doi [Astronomy \& Astrophysics] {10.1051/0004-6361/201834878}, 633, A69

\bibitem[\protect\citeauthoryear{Hill, McDonough, Toomey  \& Alexander}{Hill et~al.}{2020}]{PRD.102.043507}
Hill J.~C.,  McDonough E.,  Toomey M.~W.,   Alexander S.,  2020, \mn@doi [Phys. Rev. D] {10.1103/PhysRevD.102.043507}, 102, 043507

\bibitem[\protect\citeauthoryear{Hu}{Hu}{1998}]{Hu_1998}
Hu W.,  1998, \mn@doi [The Astrophysical Journal] {10.1086/306274}, 506, 485

\bibitem[\protect\citeauthoryear{Karwal \& Kamionkowski}{Karwal \& Kamionkowski}{2016}]{PRD.94.103523}
Karwal T.,  Kamionkowski M.,  2016, \mn@doi [Phys. Rev. D] {10.1103/PhysRevD.94.103523}, 94, 103523

\bibitem[\protect\citeauthoryear{Kazantzidis \& Perivolaropoulos}{Kazantzidis \& Perivolaropoulos}{2018}]{PRD.97.103503}
Kazantzidis L.,  Perivolaropoulos L.,  2018, \mn@doi [Phys. Rev. D] {10.1103/PhysRevD.97.103503}, 97, 103503

\bibitem[\protect\citeauthoryear{Klaewer \& Palti}{Klaewer \& Palti}{2017}]{Klaewer_2017}
Klaewer D.,  Palti E.,  2017, \mn@doi [Journal of High Energy Physics] {10.1007/jhep01(2017)088}, 2017

\bibitem[\protect\citeauthoryear{Koyama, Maartens  \& Song}{Koyama et~al.}{2009}]{Kazuya_Koyama_2009}
Koyama K.,  Maartens R.,   Song Y.-S.,  2009, \mn@doi [Journal of Cosmology and Astroparticle Physics] {10.1088/1475-7516/2009/10/017}, 2009, 017

\bibitem[\protect\citeauthoryear{Lesgourgues}{Lesgourgues}{2011}]{1104.2932}
Lesgourgues J.,  2011, The Cosmic Linear Anisotropy Solving System (CLASS) I: Overview (\mn@eprint {arXiv} {1104.2932})

\bibitem[\protect\citeauthoryear{Lewis}{Lewis}{2019}]{lewis2019getdist}
Lewis A.,  2019, GetDist: a Python package for analysing Monte Carlo samples (\mn@eprint {arXiv} {1910.13970})

\bibitem[\protect\citeauthoryear{Li \& Shafieloo}{Li \& Shafieloo}{2019}]{Li_2019}
Li X.,  Shafieloo A.,  2019, \mn@doi [The Astrophysical Journal Letters] {10.3847/2041-8213/ab3e09}, 883, L3

\bibitem[\protect\citeauthoryear{Lin, Benevento, Hu  \& Raveri}{Lin et~al.}{2019}]{PRD.100.063542}
Lin M.-X.,  Benevento G.,  Hu W.,   Raveri M.,  2019, \mn@doi [Phys. Rev. D] {10.1103/PhysRevD.100.063542}, 100, 063542

\bibitem[\protect\citeauthoryear{Lin, McDonough, Hill  \& Hu}{Lin et~al.}{2023}]{PRD.107.103523}
Lin M.-X.,  McDonough E.,  Hill J.~C.,   Hu W.,  2023, \mn@doi [Phys. Rev. D] {10.1103/PhysRevD.107.103523}, 107, 103523

\bibitem[\protect\citeauthoryear{Liu, Gao, Han, Mu  \& Xu}{Liu et~al.}{2023a}]{Liu:2023haw}
Liu G.,  Gao J.,  Han Y.,  Mu Y.,   Xu L.,  2023a, {Mitigating Cosmological Tensions via Momentum-Coupled Dark Sector Model} (\mn@eprint {arXiv} {2310.09798})

\bibitem[\protect\citeauthoryear{Liu, Mu, Gao, Han  \& Xu}{Liu et~al.}{2023b}]{Liu:2023rvo}
Liu G.,  Mu Y.,  Gao J.,  Han Y.,   Xu L.,  2023b, {The Yukawa-Coupled Dark Sector Model and Cosmological Tensions} (\mn@eprint {arXiv} {2312.01410})

\bibitem[\protect\citeauthoryear{Liu, Zhou, Mu  \& Xu}{Liu et~al.}{2023c}]{liu2023alleviating}
Liu G.,  Zhou Z.,  Mu Y.,   Xu L.,  2023c, \mn@doi [Phys. Rev. D] {10.1103/PhysRevD.108.083523}, 108, 083523

\bibitem[\protect\citeauthoryear{Liu, Mu, Zhou  \& Xu}{Liu et~al.}{2023d}]{liu2023cosmological}
Liu G.,  Mu Y.,  Zhou Z.,   Xu L.,  2023d, \mn@doi [Phys. Rev. D] {10.1103/PhysRevD.108.123546}, 108, 123546

\bibitem[\protect\citeauthoryear{Macaulay, Wehus  \& Eriksen}{Macaulay et~al.}{2013}]{PRL.111.161301}
Macaulay E.,  Wehus I.~K.,   Eriksen H.~K.,  2013, \mn@doi [Phys. Rev. Lett.] {10.1103/PhysRevLett.111.161301}, 111, 161301

\bibitem[\protect\citeauthoryear{McDonough, Lin, Hill, Hu  \& Zhou}{McDonough et~al.}{2022}]{PRD.106.043525}
McDonough E.,  Lin M.-X.,  Hill J.~C.,  Hu W.,   Zhou S.,  2022, \mn@doi [Phys. Rev. D] {10.1103/PhysRevD.106.043525}, 106, 043525

\bibitem[\protect\citeauthoryear{Mukherjee \& Banerjee}{Mukherjee \& Banerjee}{2016}]{Mukherjee2016InSO}
Mukherjee A.,  Banerjee N.,  2016, Classical and Quantum Gravity, 34

\bibitem[\protect\citeauthoryear{Mörtsell, Goobar, Johansson  \& Dhawan}{Mörtsell et~al.}{2022}]{Mortsell_2022}
Mörtsell E.,  Goobar A.,  Johansson J.,   Dhawan S.,  2022, \mn@doi [The Astrophysical Journal] {10.3847/1538-4357/ac756e}, 933, 212

\bibitem[\protect\citeauthoryear{Olivares, Atrio-Barandela  \& Pav\'on}{Olivares et~al.}{2008}]{PhysRevD.77.063513}
Olivares G.,  Atrio-Barandela F.,   Pav\'on D.,  2008, \mn@doi [Phys. Rev. D] {10.1103/PhysRevD.77.063513}, 77, 063513

\bibitem[\protect\citeauthoryear{Ooguri \& Vafa}{Ooguri \& Vafa}{2007}]{Ooguri_2007}
Ooguri H.,  Vafa C.,  2007, \mn@doi [Nuclear Physics B] {10.1016/j.nuclphysb.2006.10.033}, 766, 21

\bibitem[\protect\citeauthoryear{Palti}{Palti}{2019}]{Palti}
Palti E.,  2019, \mn@doi [Fortschritte der Physik] {10.1002/prop.201900037}, 67

\bibitem[\protect\citeauthoryear{{Planck Collaboration}, {Aghanim, N.}, {Akrami, Y.}  et~al.}{{Planck Collaboration} et~al.}{2020}]{planck2020}
{Planck Collaboration} {Aghanim, N.} {Akrami, Y.}  et~al., 2020, \mn@doi [Astronomy and Astrophysics] {10.1051/0004-6361/201833910}, 641, A6

\bibitem[\protect\citeauthoryear{Poulin, Smith, Grin, Karwal  \& Kamionkowski}{Poulin et~al.}{2018}]{PRD.98.083525}
Poulin V.,  Smith T.~L.,  Grin D.,  Karwal T.,   Kamionkowski M.,  2018, \mn@doi [Phys. Rev. D] {10.1103/PhysRevD.98.083525}, 98, 083525

\bibitem[\protect\citeauthoryear{Poulin, Smith, Karwal  \& Kamionkowski}{Poulin et~al.}{2019}]{PRL.122.221301}
Poulin V.,  Smith T.~L.,  Karwal T.,   Kamionkowski M.,  2019, \mn@doi [Phys. Rev. Lett.] {10.1103/PhysRevLett.122.221301}, 122, 221301

\bibitem[\protect\citeauthoryear{Poulin, Smith  \& Karwal}{Poulin et~al.}{2023}]{poulin2023ups}
Poulin V.,  Smith T.~L.,   Karwal T.,  2023, The Ups and Downs of Early Dark Energy solutions to the Hubble tension: a review of models, hints and constraints circa 2023 (\mn@eprint {arXiv} {2302.09032})

\bibitem[\protect\citeauthoryear{Raveri}{Raveri}{2020}]{PRD.101.083524}
Raveri M.,  2020, \mn@doi [Phys. Rev. D] {10.1103/PhysRevD.101.083524}, 101, 083524

\bibitem[\protect\citeauthoryear{Raveri \& Hu}{Raveri \& Hu}{2019}]{PhysRevD.99.043506}
Raveri M.,  Hu W.,  2019, \mn@doi [Phys. Rev. D] {10.1103/PhysRevD.99.043506}, 99, 043506

\bibitem[\protect\citeauthoryear{Riess et~al.,}{Riess et~al.}{2022}]{Riess_2022}
Riess A.~G.,  et~al., 2022, \mn@doi [The Astrophysical Journal Letters] {10.3847/2041-8213/ac5c5b}, 934, L7

\bibitem[\protect\citeauthoryear{Ross, Samushia, Howlett  et~al.}{Ross et~al.}{2015}]{stv154}
Ross A.~J.,  Samushia L.,  Howlett C.,   et~al., 2015, \mn@doi [Monthly Notices of the Royal Astronomical Society] {10.1093/mnras/stv154}, 449, 835

\bibitem[\protect\citeauthoryear{Schöneberg, Abellán, Sánchez, Witte, Poulin  \& Lesgourgues}{Schöneberg et~al.}{2022}]{SCHONEBERG20221}
Schöneberg N.,  Abellán G.~F.,  Sánchez A.~P.,  Witte S.~J.,  Poulin V.,   Lesgourgues J.,  2022, \mn@doi [Physics Reports] {https://doi.org/10.1016/j.physrep.2022.07.001}, 984, 1

\bibitem[\protect\citeauthoryear{Scolnic, Jones, Rest, Pan  et~al.}{Scolnic et~al.}{2018}]{Scolnic_2018}
Scolnic D.~M.,  Jones D.~O.,  Rest A.,  Pan Y.~C.,   et~al., 2018, \mn@doi [The Astrophysical Journal] {10.3847/1538-4357/aab9bb}, 859, 101

\bibitem[\protect\citeauthoryear{Smith, Poulin  \& Amin}{Smith et~al.}{2020}]{PRD.101.063523}
Smith T.~L.,  Poulin V.,   Amin M.~A.,  2020, \mn@doi [Phys. Rev. D] {10.1103/PhysRevD.101.063523}, 101, 063523

\bibitem[\protect\citeauthoryear{Smith, Poulin, Bernal, Boddy, Kamionkowski  \& Murgia}{Smith et~al.}{2021}]{PRD.103.123542}
Smith T.~L.,  Poulin V.,  Bernal J.~L.,  Boddy K.~K.,  Kamionkowski M.,   Murgia R.,  2021, \mn@doi [Phys. Rev. D] {10.1103/PhysRevD.103.123542}, 103, 123542

\bibitem[\protect\citeauthoryear{T\'ellez-Tovar, Matos  \& V\'azquez}{T\'ellez-Tovar et~al.}{2022}]{PRD.106.123501}
T\'ellez-Tovar L.~O.,  Matos T.,   V\'azquez J.~A.,  2022, \mn@doi [Phys. Rev. D] {10.1103/PhysRevD.106.123501}, 106, 123501

\bibitem[\protect\citeauthoryear{Ureña-López \& Gonzalez-Morales}{Ureña-López \& Gonzalez-Morales}{2016}]{Ure_a_L_pez_2016}
Ureña-López L.~A.,  Gonzalez-Morales A.~X.,  2016, \mn@doi [Journal of Cosmology and Astroparticle Physics] {10.1088/1475-7516/2016/07/048}, 2016, 048

\bibitem[\protect\citeauthoryear{Verde, Treu  \& Riess}{Verde et~al.}{2019}]{Verde_2019}
Verde L.,  Treu T.,   Riess A.~G.,  2019, \mn@doi [Nature Astronomy] {10.1038/s41550-019-0902-0}, 3, 891

\bibitem[\protect\citeauthoryear{Wang, Abdalla, Atrio-Barandela  \& Pavón}{Wang et~al.}{2016}]{Wang_2016}
Wang B.,  Abdalla E.,  Atrio-Barandela F.,   Pavón D.,  2016, \mn@doi [Reports on Progress in Physics] {10.1088/0034-4885/79/9/096901}, 79, 096901

\bibitem[\protect\citeauthoryear{Wojtak \& Hjorth}{Wojtak \& Hjorth}{2022}]{stac1878}
Wojtak R.,  Hjorth J.,  2022, \mn@doi [Monthly Notices of the Royal Astronomical Society] {10.1093/mnras/stac1878}, 515, 2790

\bibitem[\protect\citeauthoryear{Yang, Pan  \& Mota}{Yang et~al.}{2017}]{PhysRevD.96.123508}
Yang W.,  Pan S.,   Mota D.~F.,  2017, \mn@doi [Phys. Rev. D] {10.1103/PhysRevD.96.123508}, 96, 123508

\bibitem[\protect\citeauthoryear{Zhou, Liu, Mu  \& Xu}{Zhou et~al.}{2022}]{Zhouzh}
Zhou Z.,  Liu G.,  Mu Y.,   Xu L.,  2022, \mn@doi [Monthly Notices of the Royal Astronomical Society] {10.1093/mnras/stac053}, 511, 595

\makeatother
\end{thebibliography}




\appendix
\section{The Full MCMC posteriors}

\begin{figure*}
	\includegraphics[width=\linewidth]{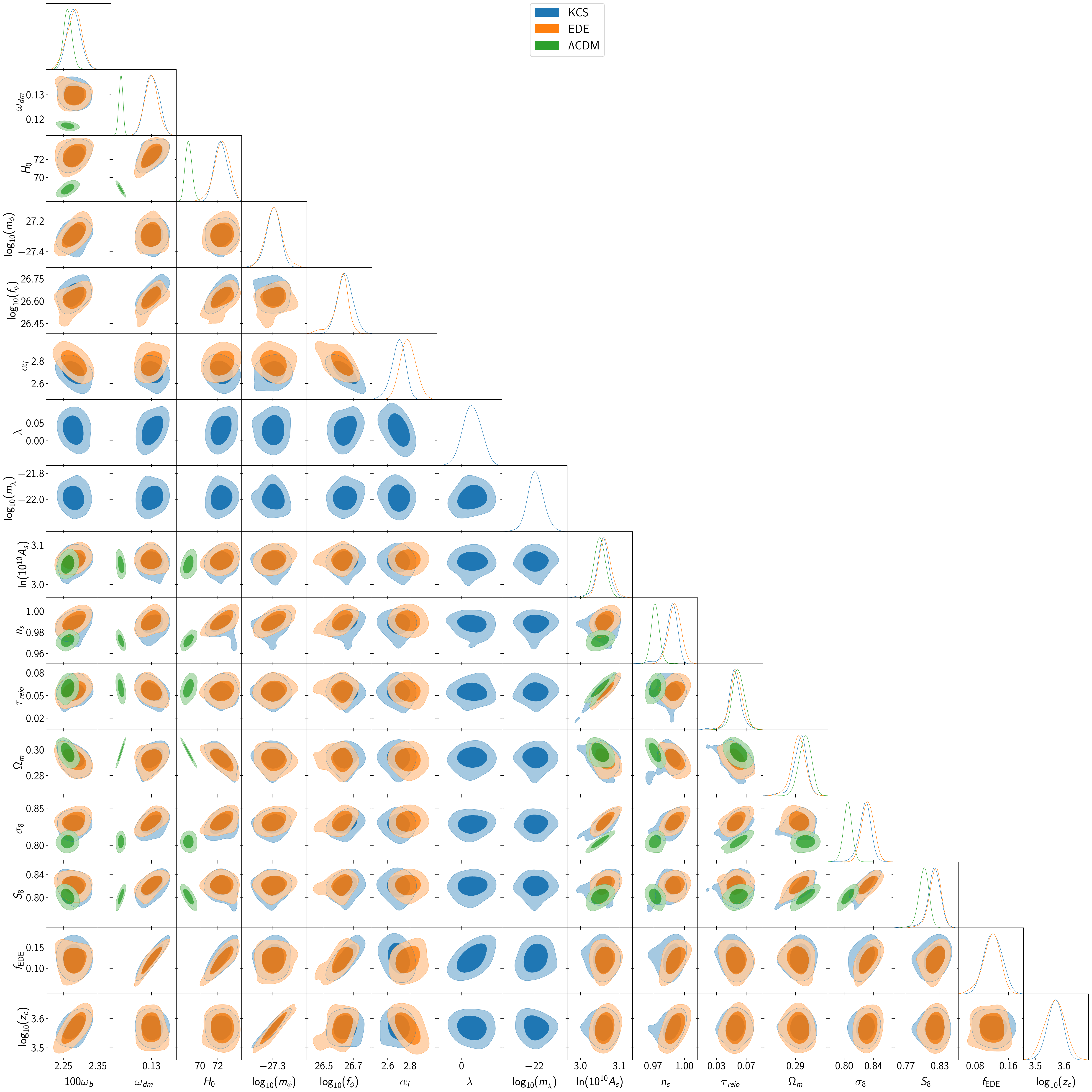}
    \caption{
		The complete posterior distributions for $\Lambda$CDM, EDE, and KCS models 
        with data comprising CMB, BAO, SNIa, SH0ES, and $S_8$ from DES-Y3. }
    \label{fig:8}
\end{figure*}



\bsp	
\label{lastpage}
\end{document}